\documentclass[12pt]{article}
\usepackage[utf8]{inputenc} 
\usepackage[T1]{fontenc}
\usepackage{amssymb,amsmath}
\usepackage[english]{babel}
\usepackage[utf8]{inputenc}
\usepackage{graphicx} 
\usepackage{fancyhdr}
\usepackage{amsmath}
\usepackage{physics}
\usepackage{slashed}
\usepackage{cite}
\usepackage{hyperref}

\usepackage{stackengine}
\usepackage{mattens}
\usepackage{eucal}
\usepackage[usenames, dvipsnames]{color}
\numberwithin{equation}{section}
\usepackage{cite}
\newcommand{\bb}{\begin{bf}} 
\newcommand{\eb}{\end{bf}}

\newcommand{\ben}{\begin{enumerate}} 
\newcommand{\een}{\end{enumerate}} 
\newcommand{\bi}{\begin{item}} 
\newcommand{\ei}{\end{item}} 
\newcommand{\bc}{\begin{center}} 
\newcommand{\ec}{\end{center}} 
\newcommand{\bl}{\begin{Large}} 
\newcommand{\el}{\end{Large}} 

\newcommand{\beq}{\begin{equation}}  
\newcommand{\eeq}{\end{equation}}  
\newcommand{\bea}{\begin{eqnarray}}  
\newcommand{\eea}{\end{eqnarray}}

\def\b{$\beta$}

\def\i{$\iota$}

\newcommand\iso[2]{\mbox{${}^{#2}${\rm #1}}}

\def\b1#1{\iso{B}{1#1}}

\begin{document} 
\begin{titlepage} 
\begin{flushright}
 \end{flushright} 

\begin{center} 
\vspace*{1.5cm} 
 
{\Large{\textbf {{Gravitational Waves  and Primordial Black Holes from \\[2mm]
   Supersymmetric  Hybrid Inflation }}}}\\ 
 \vspace*{10mm} 
\end{center} 

\begin{center}
 {\bf Vassilis~C.~Spanos and Ioanna~D.~Stamou }
\vspace{.7cm}

{\it National and Kapodistrian University of Athens, Department of Physics, \\
 Section of Nuclear {\rm \&} Particle Physics,  GR--15784 Athens, Greece} \\

\end{center}

\vspace{1.0cm}

\begin{abstract}
We  study the effect of   supergravity  corrections due 
  to  a linear and a squared  term  in the K\"ahler potential,  in the context of
 a supersymmetric hybrid inflation model. By appropriate choice of the 
parameters  associated to these terms, we are able to satisfy the main cosmological  
constraints for 
the spectral index $n_s$ and the tensor-to-scalar ratio $r$. In addition,  this model 
predicts primordial black hole abundance enough to account for the whole dark matter of the Universe 
and gravitational wave spectra within the reach of future detection experiments. 
The predictions of the model can be made compatible to the  NANOGrav reported signal, at the cost of  
significantly lower primordial black hole abundance. 
\end{abstract}

\end{titlepage} 

\baselineskip= 15.2 pt

\section{Introduction}

An important milestone in  Cosmology was the  detection of Gravitational Waves (GWs) by LIGO and
Virgo collaborations~\cite{Abbott:2016blz,Abbott:2016nmj,Abbott:2017oio}.  The
detection of such  signals, related to the merging of black holes or neutron stars, 
triggered numerous  studies, exploring  the possibility that the primordial black holes (PBHs)
constitute a significant part  or the whole  of the Dark Matter (DM) of the Universe. In addition,
recently, it was reported   strong evidence for stochastic common-spectrum process  by the NANOGrav collaboration~\cite{Arzoumanian:2020vkk,Arzoumanian:2018saf,Aggarwal:2018mgp}. 
Needless to say, that analogous and more precise detection  signals for  GWs 
are expected 
from future space-based   GW interferometers such as  LISA,  BBO, DECIGO, SKA and Tianquin,\cite{Audley:2017drz,Sato:2017dkf,Sathyaprakash:2009xs,Yagi:2011wg,Luo:2015ght,Zhao:2013bba}.

As a result, many theoretical studies have appeared  in the literature~\cite{Ballesteros:2017fsr,Gao:2018pvq,Cicoli:2018asa,Dalianis:2018frf,Garcia-Bellido:2017mdw,Ezquiaga:2017fvi,Nanopoulos:2020nnh,Stamou:2021qdk,Hertzberg:2017dkh,Ballesteros:2019hus,Kefala:2020xsx,Dalianis:2021iig,Braglia:2020fms,Braglia:2020eai,Braglia:2020taf,Fumagalli:2020adf,Palma:2020ejf,Gundhi:2020kzm,Khlopov:2015oda,Ketov:2019mfc,Mukherjee:2021ags,Mukherjee:2021itf,Vagnozzi:2020gtf,Pi:2017gih,Cai:2019bmk,Kohri:2020qqd,Ashoorioon:2019xqc,Ashoorioon:2020hln},  which explain the production of PBH as a fraction of DM of the Universe. This PBH production during the radiation dominance 
epoch,  can be associated  to  a significant enhancement of the scalar power spectrum. Many of these models are based on single field inflation   and especially on models with  a near inflection point in the effective scalar potential~\cite{Ballesteros:2017fsr,Gao:2018pvq,Cicoli:2018asa,Dalianis:2018frf,Garcia-Bellido:2017mdw,Ezquiaga:2017fvi,Nanopoulos:2020nnh,Stamou:2021qdk,Hertzberg:2017dkh,Ballesteros:2019hus}. 
The required enhancement factor of the power spectrum is calculated to be seven  order of
magnitudes~\cite{Ballesteros:2017fsr,Gao:2018pvq,Cicoli:2018asa,Dalianis:2018frf,Garcia-Bellido:2017mdw,Ezquiaga:2017fvi,Nanopoulos:2020nnh,Stamou:2021qdk,Hertzberg:2017dkh,Ballesteros:2019hus}.  A severe   drawback of these models,  is the  high level of the necessary fine-tuning,  in order to achieve such a big amplification of the power spectrum. For this reason, alternative methods have been proposed  to alleviate the issue of the fine-tuning, for example the  multi-field inflation  models~\cite{Braglia:2020fms,Braglia:2020eai,Braglia:2020taf,Fumagalli:2020adf,Palma:2020ejf,Gundhi:2020kzm} or models with a step behavior  in the effective scalar potential\cite{Adams:2001vc,Hazra:2010ve,Kefala:2020xsx,Dalianis:2021iig}.

In this paper   we follow a different avenue  in order to explain the generation of GWs and the production of PBHs.  
Specifically, we study the case of a hybrid inflation  model~\cite{Linde:1993cn,Linde:1991km}, including supergravity (SUGRA) corrections.
The main advantage of the hybrid models, is that the required fine-tuning is significantly 
smaller than in models where the PBHs
are produced due to an  inflection point in the scalar potential. 
The characteristic feature of a  hybrid model, is that the minimum of the scalar effective  potential
corresponds to the false vacuum with non-vanishing  energy density \cite{Linde:1993cn,Copeland:1994vg}. 
This false vacuum dominates and  becomes unstable, when the inflaton field acquires a critical value\cite{Copeland:1994vg}.
Needless to say,  that the inflationary  predictions from the hybrid false vacuum  models are quite
different than the true vacuum models.  
Specifically, in the true vacuum models  inflation ends with oscillations in the minimum of the potential,
which designate the reheating. 
Moreover, it appears that the inflation scale in true vacuum models differs from this in  the false vacuum. 
In particular,   in the false vacuum models  inflation occurs far below the Planck scale, therefore  they are 
similar  to the  embedded   SUGRA  models~\cite{Copeland:1994vg}. 
On the contrary,  the inflation scale in the true vacuum models is predicted to be usually  
 high.

The main drawback of hybrid inflationary models was  their prediction 
for the spectral index $n_s \gtrsim 1$  ~\cite{Copeland:1994vg,Mollerach:1993sy,Linde:1997sj},
value excluded by the Planck data~\cite{Akrami:2018odb,Ade:2015lrj}.   
In~\cite{Dvali:1994ms,Bastero-Gil:2006zpr} introduced one-loop radiative
corrections in order to achieve a reduction to $n_s$, although during the last years 
alternative methods  have been proposed~\cite{Dimopoulos:1997fv,Dvali:1997uq,Panagiotakopoulos:1997if,Tetradis:1997kp,Lazarides:1998zf,Clesse:2010iz,Pallis:2009pq,Armillis:2012bs,Pallis:2013dxa,Pallis:2014xva,Dimopoulos:2016tzn,Kadota:2017dbz,Rehman:2009nq,Yamaguchi:2004tn}.
Specifically, in the context of  no-scale SUGRA  hybrid models,  a reduced 
value for $n_s$ can be predicted~\cite{Wu:2016fzp,Moursy:2020sit,Antusch:2009ef}.
 In the context of hybrid models the  production of PBHs has already be studied~\cite{Lyth:2011kj,Clesse:2015wea,Kanazawa:2000ea,Choi:2021yxz}, 
also with small values  for $n_s$. This PBH production obtained by an appropriate enhancement of the power spectrum of the scalar perturbations. In addition, the production of GWs can be obtained by this enhancement~\cite{Garcia-Bellido:2007bcw,Civiletti:2011qg,Kawasaki:2012rw,Lazarides:2020zof,Lazarides:2015cda,Cai:2018dig,Cai:2019amo,Domenech:2021wkk,Pi:2020otn}.

In this work we present a pure hybrid model, 
 where we achieve cosmologically accepted values for $n_s<1.0$ due to SUGRA corrections. 
These corrections are related to a linear and a squared  term, added in the K\"ahler potential.
The enhancement of the power spectrum in hybrid models, such as  shown in Ref.~\cite{Clesse:2015wea}, occurs during the mild waterfall phase. 
Specifically, these models are based on a two-field inflation scenario, where one field  acquires tachyonic solution in the critical point and the other, which plays the role of the inflaton, becomes unstable. Thus, the analysis of the background dynamics is decomposed in two phases: a slow-roll phase until the critical point and a second waterfall phase 
till the end of the inflation. The evaluation of the inflationary observables during these phases has been performed 
analytically in the slow-roll approximation~\cite{Clesse:2015wea,Kodama:2011vs}. For comparison 
we have performed also a  numerical solution of  the background equations as well as 
the  perturbations.

In the context of our model we have chose two representative 
sets of  parameters  for the extra K\"ahler terms.
In any case  we are able to satisfy the main cosmological  constraints for 
the spectral index $n_s$ and the tensor-to-scalar ratio $r$. 
Moreover,  this model 
predicts PBH abundance enough to account for the total DM of the Universe 
and GW spectra that can be detected  in the future   experiments like LISA, DELIGO etc. 
With an appropriate choice of the parameters,  the GW spectrum 
 can be made compatible to  NANOGrav signal, but in this case the 
 PBH  abundance is  much smaller,  due to the SUBARU constraints. 

The layout of the paper is as follows: In Section 2 we introduce the hybrid model with the SUGRA
 correction as discussed before,  in order to derive acceptable values for the  spectral index $n_s$ and the
  tensor-to-scalar ratio $r$. In Section  3 we present the background dynamics of the waterfall as well as the  perturbation of the fields. 
  We calculate analytically the scalar power spectrum based on the slow-roll approximation  and for comparison we evaluate 
  also the spectrum using  a numerical solution. 
  In Section 4 we evaluate the energy density of the GWs and we show that the 
  model yields detectable spectra for the future experiments, like LISA, 
  but also can explain the reported signal from the NANOGrav collaboration. 
  In Section 5 
  we evaluate the fractional abundance of PBHs,  which   for a particular choice of the parameters, 
  can  account for  the whole DM of the Universe.
  Furthermore, in Section 6 we estimate the amount  of the required  fine-tuning 
  and we  present  our concluding remarks.

\section{The Hybrid Model}
\label{sect:model}

The  hybrid model~\cite{Linde:1993cn,Copeland:1994vg,Dvali:1994ms} results  from    the globally supersymmetric
renormalizable superpotential
\begin{equation}
W= \kappa \, S  (  \Psi_1 {\Psi}_2  - m^2)\, ,
\label{spotential}
\end{equation}
where $ \Psi_1, \Psi_2$   are chiral superfields,
 the scalar component of the 
superfield,  $S$ is the gauge singlet  inflaton field, $\kappa$ is a dimensionless  coupling constant and $m$ is a mass.  It is important to notice that this   superpotential is symmetric under $R$-symmetry\footnote{{For the charges of the superfields in the global R-symmetry we consider that the field $S$ has $U(1)_R$ equals to 1 and the fields $\Psi_1,\Psi_2$ has  $U(1)_R$ equals to 0 \cite{Lazarides:1998zf}.}}, or in other words, transformations of the field such as $X \rightarrow  e^{i \omega} X$ lead to the transformation of the superpotential $W \rightarrow e^{i \omega}W$. This symmetry removes the undesirable self-couplings of  the inflaton field $S$,  and it is the only  symmetry which treats the false vacuum in a natural way~\cite{Dvali:1994ms}.

The scalar potential in SUSY is given from the following expression

\begin{equation}
{V_F^{SUSY}= \sum_i \left| \frac{\partial W}{\partial F_i}\right|^2}
\end{equation}
{where $F_i$ indicates the superfields: $\Psi_1,\Psi_2$ and $S$. In the case of superpotential (\ref{spotential}), one gets we have:}
\begin{equation}
{V_F^{SUSY}= \left|\kappa(\Psi_1{\Psi}_2-m^2) \right|^2 +  \left| \kappa S\Psi_1\right|^2 + \left|\kappa S {\Psi}_2 \right|^2}
\end{equation}

{We consider the following form of the K\"ahler potential: }
\begin{equation}
K=S\bar{S}+   \Psi_1{\bar{\Psi}_1}+ \Psi_2{\bar{\Psi}_2}
\label{eq:firstkahler_susy}
\end{equation}
{ For the choice of Eq.~(\ref{eq:firstkahler_susy}) we have the following K\"ahler metric}
\begin{equation}
{K_{i\bar{j}} =\frac{\partial^2K}{\partial F^i\partial{F}^{\bar j}}=\begin{pmatrix}
1 &0 & 0\\
0& 1 & 0\\
0& 0& 1 \\
\end{pmatrix}.}
\label{eq:pinakas_kahler_epipleon_eksigisi}
\end{equation} 
{So the kinetic term of the Langrangian is given}
\begin{equation}
{L_{kin}={K_{i\bar{j}}\partial_{\mu}F^i\partial^{\mu} F^{\bar j}}}
\end{equation}
Hence the scalar potential is given from
\begin{equation}
{  V_F^{\text{SUSY}}= \Lambda\left[ \left(1-\frac{\psi^2}{M^2}\right)^2+ \frac{2 \phi^2 \psi^2}{M^4 }\right]  }\, , 
\label{firstsusy}
\end{equation} 
where we have assumed {$\Lambda=\kappa^2 m^4$ and  $ M^2 \equiv2m^2$}. {  In order to fix the non-canonical kinetic term we have $|S|=\phi /\sqrt{2}$ and $|\Psi_1|=| \Psi_2|=\psi/\sqrt{2}$ ~\cite{Copeland:1994vg,Lazarides:1998zf}}. 
The potential is flat along the direction $\psi=0$,  $|\phi|>|\phi_c|=m\sqrt{2}$ and  is given by a constant value for the energy density  $V=\kappa^2 m^4$. One the other hand, the field $\psi$ develops  tachyonic solutions if 
\begin{equation}
{\kappa^2(-2m^2+\phi^2+3\psi^2)<0 }\, , 
\end{equation}
Along the flat direction this  condition  becomes,
\begin{equation}
{\phi^2<\phi_c^2=2m^2 \equiv M^2}\, ,
\label{fcrit}
\end{equation}
where $\phi_c$ is the critical value of the field $\phi$, after which this field $\psi$ becomes tachyonic. 
%
The main  phenomenological issue of  this  model is that the predicted  spectral index, $n_s$, is  almost equal to unity, value  rejected by the  Planck  data~\cite{Akrami:2018odb}. 
It has been proposed that incorporating the one-loop radiative corrections,   leads to a red spectral index~\cite{Dvali:1994ms}. 

However, there is an alternative  way to fix  this problematic feature. 
By  incorporating    SUGRA corrections,  one maintains the basic  characteristics  of the model, such as
the waterfall behavior or the false vacuum inflation,  and predicts  $n_s \lesssim  1$. 
Specifically we consider a  K\"ahler potential like
\begin{equation}
    K= S \bar{S} +b_1(S +\bar{S}) +b_2(S +\bar{S})^2  + \Psi_1{\bar{\Psi}_1}+ \Psi_2{\bar{\Psi}_2}\,
\label{eqkahler}
\end{equation}
where $b_{1}$ is a dimensionful parameter with mass dimensions and $b_2$ is dimensionless. 
As for the origin of these parameters here we will remain agnostic, but
their role and their phenomenologically   preferred   values 
will be discussed in the following. {This form of K\"ahler potential is similar to Eq.~(\ref{eq:firstkahler_susy}) plus a shift-symmetric term~ \cite{Ketov:2016gej}}. 

We calculate the $F$-term of  the scalar potential using its  general form is SUGRA models~\cite{Cremmer1979}
\begin{equation}
   V_F^{\text{SUGRA}}= e^{K/{M_P}^2}\left[\left( {K^{-1}}\right)^i_{\bar{j}}
              \left(W^{\bar{j}}+\frac{WK^{\bar{j}}}{{M_P}^2}\right) \left(\bar{W}_i+\frac{\bar{W}K_{i}}{{M_P}^2}\right)- \frac{3|W|^2}{{M_P}^2} \right]  \, , 
   \label{eqvgeneral}
\end{equation}
where $({K^{-1}})^i_{\bar{j}}$ is the inverse K\"ahler metric,{ $K_i=\frac{\partial K}{\partial F^i}$ and $W_i= \frac{\partial W}{\partial F^i}$}. $M_P$ is the reduced Planck mass\footnote{In the following we keep 
$M_P$ as dimensionful parameter, in order to  clarify  the SUGRA limiting behavior. }. 
 The indices $i,j$ in this equation and thereafter, run over the chiral superfields   $S$, $\Psi_1$ and $\Psi_2$  and with bar we denote the conjugate pair of each superfield. 
For the superpotential, we use  the~(\ref{spotential}). %
In order to calculate the  corrections in the effective scalar potential 
we expand   the exponential in Eq.~(\ref{eqvgeneral}) as
\begin{equation}
{e^{K/M_P^2}=1+\frac{K}{ M_P^2}+\frac{K^2}{2\, M_P^4} + \mathcal{O} \left(\frac{1}{ M_P^6}\right) \, .}
\label{eq:expon} 
\end{equation}
The K\"ahler metric for  the K\"ahler potential (\ref{eqkahler}) takes the form 
\begin{equation}
{K_{i\bar{j}} =\begin{pmatrix}
1+2b_2 &0 & 0\\
  0& 1 & 0\\
0& 0& 1 \\
\end{pmatrix} }.
\label{eqapa3}
\end{equation} 
Using in (\ref{eqvgeneral}),  the Eqs.~(\ref{spotential}), (\ref{eqkahler}), (\ref{eq:expon}) and   (\ref{eqapa3}),    the scalar potential reads as 
\begin{equation}
{
V_F^{\text{SUGRA}}=\frac{\kappa^2(M^4-2M^2\psi^2+2\psi^2\phi^2+\psi^4)}{4+8b_2}+\frac{\mathcal{A}_1}{M_P^2}+\frac{\mathcal{A}_2}{M_P^4}+ \mathcal{O} \left( \frac{1}{M_P^6} \right)}  \, ,
\label{eq:fullpotential}
\end{equation}
{where we have assumed that $|\Psi_1|=| \Psi_2|=\psi/\sqrt{2}$, as before. For the superfield $S$, we need to fix the non-canonical kinetic term  by considering  the K\"ahler metric in Eq.~({\ref{eqapa3}). Hence we have considered the following redefinition of the field }}
\begin{equation}
S = \frac{\phi}{\sqrt{2+4b_2}} \, .
\end{equation} 
 The $\mathcal{A}_{1,2}$ are functions of $M, \phi, \psi$ with mass dimensions 6 and 8,  respectively.
Detailed expression for these can be found in Appendix \ref{sect:app} in  Eqs.~(\ref{eq:apA1}) and (\ref{eq:apA2}). 
One can notice that the first  term in  (\ref{eq:fullpotential}) is $V_F^{\text{SUSY}}/(1+2b_2)$.
In the limit  $M_P \rightarrow \infty$ and $b_2 \rightarrow 0$ apparently  one derives the SUSY potential as in {Eq.~(\ref{firstsusy})}.
\begin{figure}[t!]
\centering
\includegraphics[width=80mm]{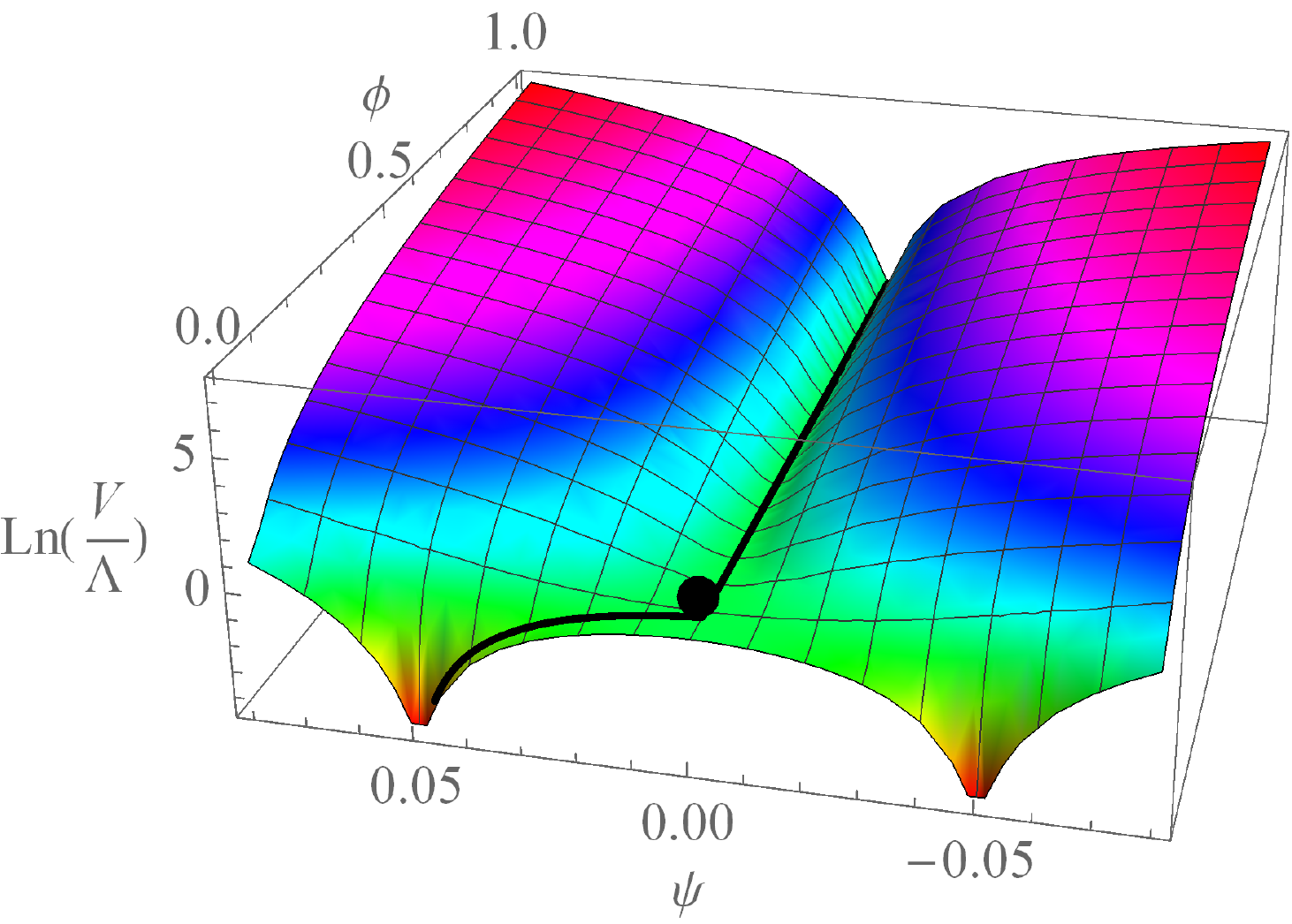}
\caption{ The quantity $\ln\left( \frac{V}{\Lambda} \right)$,  using   Eq.~(\ref{eq.potsugra}) with  $b_1=3.506\times 10^{-4}\,M_P$, 
$b_2=-3.5\times 10^{-3}$ and $m=0.05\, M_P$.  The fields $\phi$ and $\psi$ 
are measured in $M_P$ units. The black bullet denotes the position of the critical point. }
\label{f1}
\end{figure}
Therefore, the total effective scalar potential is 
\begin{equation}
        V \equiv V_F^{\text{SUGRA}}=   \Lambda\left[ \left(1-\frac{\psi^2}{M^2}\right)^2+ \frac{2 \phi^2 \psi^2}{M^4 } +\tilde{V}_F^{\text{correction}} \right],
        \label{eq.potsugra}
\end{equation}
where { $V_F^{\text{correction}}= V_F^{\text{SUGRA}} -V_F^{\text{SUSY}}$} from Eqs.~(\ref{firstsusy}) and (\ref{eq:fullpotential}) and $\tilde{V}_F^{\text{correction}}=V_F^{\text{correction}}/\Lambda$. If we expand this expression around the critical value of $\phi$ we derive\cite{Clesse:2015wea,Bernardo:2016jdr} 
\begin{equation}
   \tilde{V}_F^{\text{correction}} = a_0+a_1 (\phi- \phi_c)+ a_2 (\phi- \phi_c)^2+a_3 (\phi- \phi_c)^3+a_4 (\phi- \phi_c)^4.
\label{sugraexpan}   
\end{equation}
In this  expansion,  terms up to $a_2$ are related to $\mathcal{A}_1$ in (\ref{eq:fullpotential}), 
where these up  to  $a_4$ are related  to $\mathcal{A}_2$.
The numerical  calculation reveals that the terms which appear  
in  (\ref{eq:fullpotential}),  suffice to approximate the  $ V_F^{\text{SUGRA}} $ in (\ref{eqvgeneral}). 
More details can be found in the Appendix.

 In Fig.~\ref{f1}  we present the effective scalar potential in Eq. (\ref{eq.potsugra}). As one can notice, the inflaton field moves through the valley until it reaches the critical point. After that the other field, which is  called waterfall, acquires tachyonic solution and the inflaton moves through the waterfall. Finally, the inflation ends in false vacuum.
 We can show that the prediction of the $n_s$ in this model  is  indeed compatible with the Planck cosmological data.
In the slow-roll limit it is known that 
\begin{equation}
 n_s=1+2 \eta_V-6 \epsilon_V \, , 
 \label{ns_cal}
\end{equation}
with   
\begin{equation}
\varepsilon_V= \frac{M_P^2}{2}\left(\frac{V'}{V} \right)^2 , \quad \eta_V=M_P^2  \, \frac{V''}{V}\, . 
\end{equation}
Prime  denotes  derivation  with  respect to the  fields. 
It is worth noting  that in this model  we get  $n_s=0.965$,    consistent with the current data. 
 For the tensor-to-scalar ratio we use  the corresponding slow-roll expression
\begin{equation}
r=16 \, \varepsilon_V    \, .
\label{r_cal}
\end{equation}
Detailed phenomenological analysis delineating values for $b_{1,2}$ compatible with the experimental constraints 
for $n_s$ and $r$ will be discussed in the following section.


\section{Background Dynamics And  Perturbations Along The Waterfall}

 In the hybrid inflation  models   two scalar fields are required: the inflaton field $\phi$ and the waterfall field $\psi$. 
 The  field  $\psi$  becomes tachyonic at some critical value $\phi=\phi_c$, where its mass squared  gets 
 negative values. In this section we discuss  the  background dynamics for  both $\phi$ and  $\psi$ fields.
 In addition, we present the evaluation of scalar power spectrum with analytical and numerical tools, adopting  methods 
 from Refs~\cite{Clesse:2015wea,Kodama:2011vs}.

  \begin{table}[h!]
\centering
 \begin{tabular}{|c|c|c|c|c|c|c|}
\hline
  Model&$M(M_P)$&$b_1(10^{-4}M_P)$&$b_2(10^{-3})$ &  $\Lambda (10^{-20} M_P^4)$ &$n_s$  & $r$  \\
\hline
set 1 &$0.05$& ${3.506}$& ${-3.5} $& ${4.9}$ & ${0.9651}$ &${1.55 \times 10^{-12}}$  \\
\hline
 set 2 &$0.10$& $8.918$& $-5.0 $&  $140$ &  $0.9684$ &$1.54 \times 10^{-10}$ \\
\hline
\end{tabular}
 \caption{ The prediction for the spectral index $n_s$ and ratio tensor-to-scalar $r$ for two 
 representative sets of the parameters $b_{1,2}$.}
 \label{tabunsr}
\end{table}

\begin{figure}[t!]
\centering
\includegraphics[width=80mm]{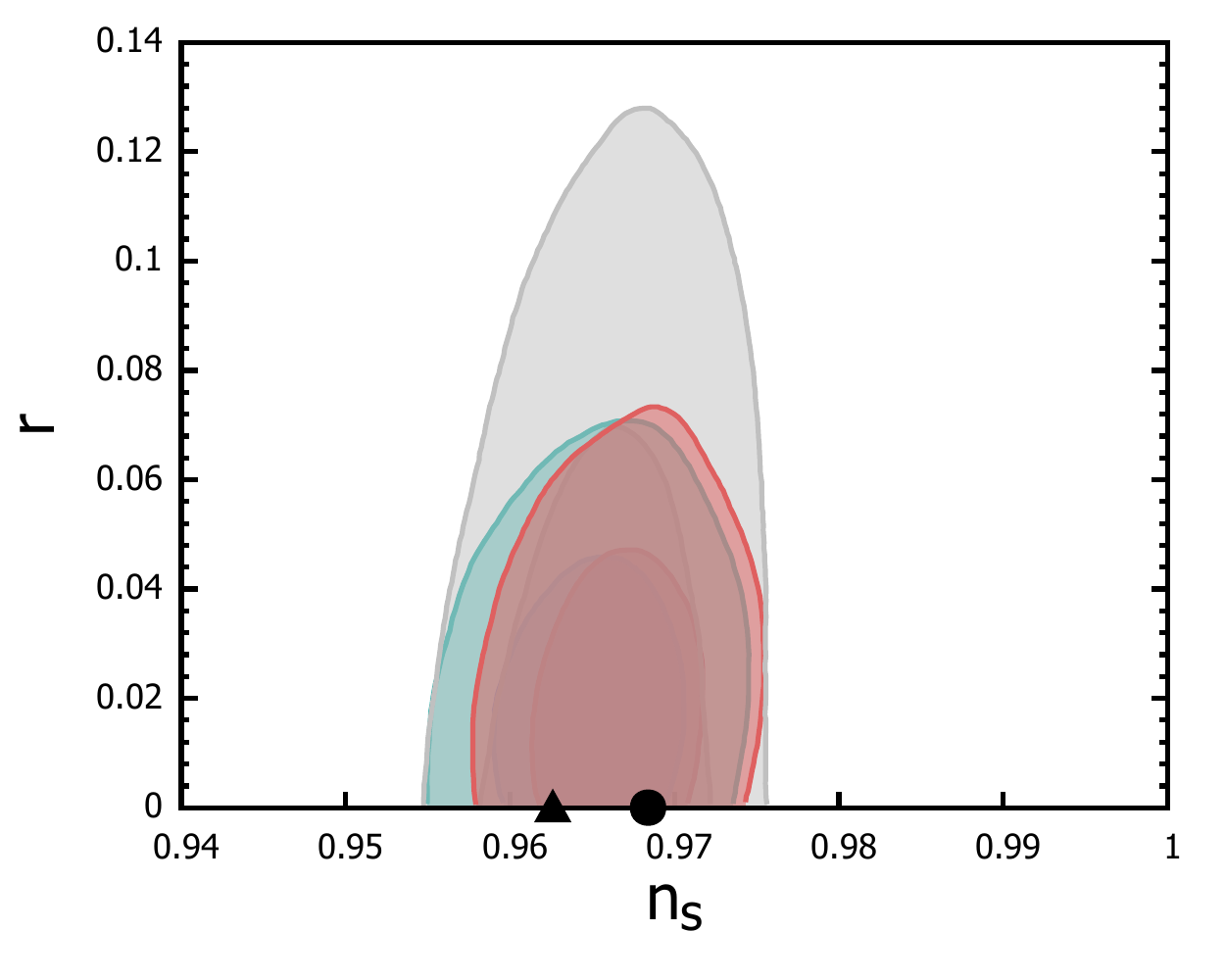}
\caption{ The predictions of our models for $n_s$ and $r$ plotted against the 
  Planck 2018 constraints~\cite{Akrami:2018odb}. 
  Triangle   corresponds to set 1 and bullet to set 2,  as given in Table \ref{tabunsr}.}
\label{fnsr} 
\end{figure}

  First we calculate  the evolution of the fields  with respect to the cosmic time, as described  from the potential~(\ref{eq.potsugra}). 
Consequently,  we evaluate the spectral index $n_s$ 
and the tensor-to-scalar ratio $r$ using the Eqs.~(\ref{ns_cal}) and (\ref{r_cal}), respectively.
Our results for two sets for the parametres $b_{1,2}$ are shown 
in Table~\ref{tabunsr}\footnote{The numerical value of the   parameter $\Lambda$ is fixed by the requirement $P_R=2.1\times 10^{-9}$ at the CMB scale. }.
We notice that the additional terms, that depend on $b_{1,2}$ are found to be numerically at least three orders of magnitude smaller 
than the dominant $M$-term. On the other hand, their presence is important in order to satisfy the cosmological constraints. 
This can be understood, because the subdominant terms contribute in the derivatives of the potential $V$ and through these to $n_s$.
In Fig.~\ref{fnsr} we plot these predictions for the observables  $n_s$ and $r$ along with the
current  allowed regions  
 by the Planck 2018 collaboration~\cite{Akrami:2018odb}.

\begin{figure}[h!]
\centering
\includegraphics[width=160mm]{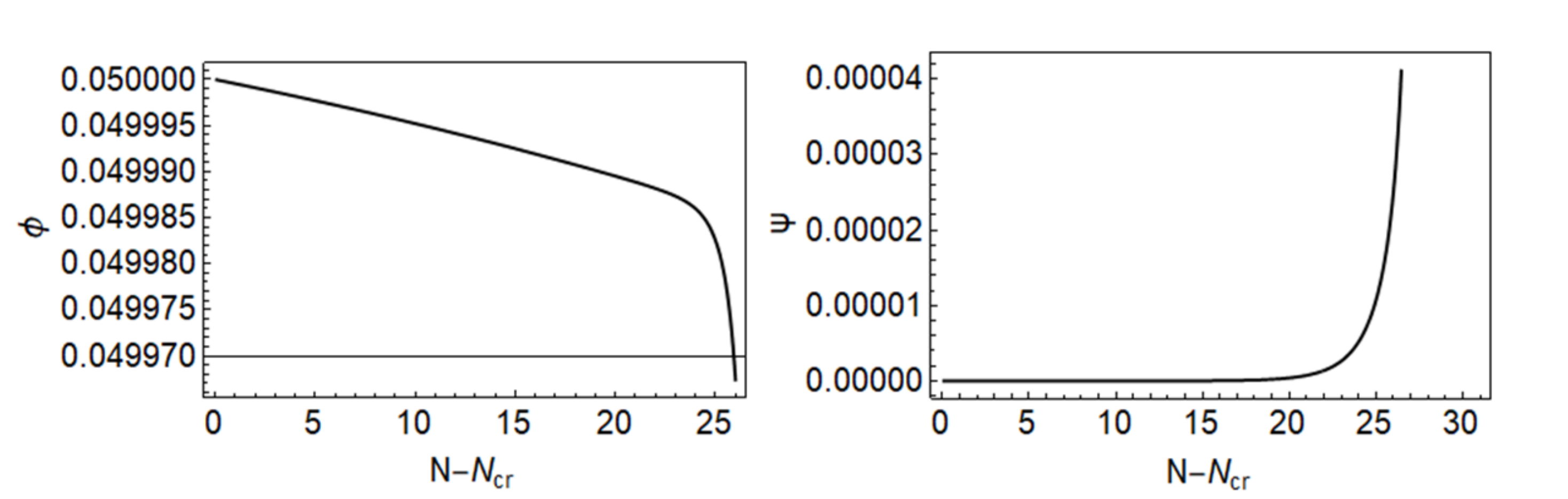}
\caption{ {The fields $\phi$, $\psi$ as  functions of 
the number of e-folds $N$ in $M_P$ units, starting from the critical point. We evaluate them   solving the full  background  Eqs.~(\ref{eq:bckgeq}) 
using the scalar potential (\ref{eq.potsugra}).}} 
\label{f5}
\end{figure}

The background dynamics is described by the Friedmann-Lema\^itre equation
\begin{equation}
H^2=\frac{1}{3M_P^2} \Big(  \frac{\dot \phi^2}{2}+\frac{\dot \psi^2}{2}+V(\phi,\psi) \Big)
\end{equation}
and the Klein-Gordon equations
\begin{equation}
\begin{split}
\ddot \phi +3H\dot \phi + \frac{\partial V}{\partial \phi}=0\\
\ddot \psi+3H\dot \psi + \frac{\partial V}{\partial \psi}=0 \, , 
\label{eq:bckgeq}
\end{split}
\end{equation}
where the dots  denote  differentiation with respect to cosmic time. In Fig.~\ref{f5} 
we display the full solution  of the (\ref{eq:bckgeq}) for the fields $\phi$ and $\psi$. 
As initial conditions for the solution we assume  those of the critical point. 
Specifically, we assume $\phi_{ic}=\phi_c=0.05\, M_P$ and $\psi_{ic}= 1.3 \times 10^{-10}\, M_P$
 and we calculate the background equations until the end of inflation, or equivalently until the parameter $\varepsilon= -\dot H / H^2$ reaches the value 1.

The potential,  we proposed in Eq. (\ref{eq.potsugra}),
 gives significant enhancement in the power spectrum, almost by seven order of magnitude, 
  due to the waterfall behavior, as it has been pointed out in~\cite{Clesse:2015wea}. In the following we present the analytical  calculation of the power spectrum based on the
    slow-roll approximation and for  comparison  the corresponding numerical result based on the  integration of 
    curvature perturbations of the fields.

 \subsection{The slow-roll approximation}
The equations of motion (\ref{eq:bckgeq}) can be solved  numerically. However, it is  possible to derive analytical solutions by considering the usual slow-roll approximation~\cite{Kodama:2011vs,Clesse:2015wea,Clesse:2013jra}. In this subsection we derive the analytical solution 
and we use these results in order to evaluate   the scalar power spectrum~\cite{Clesse:2015wea}.

The equations of motion in slow-roll approximation, ignoring the seconds derivatives in  (\ref{eq:bckgeq}),  are  
\begin{equation}
\begin{split}
3H \dot\phi= -\frac{\partial V}{\partial \phi} \\
3H \dot\psi= -\frac{\partial V}{\partial \psi}
\end{split}
\end{equation}
and the Friedmann-Lema\^itre equation is 
 \begin{equation}
 H^2=\frac{ \Lambda}{3M_P^2}.
 \end{equation}
We consider the potential as in Eq.~(\ref{eq.potsugra}) with  the SUGRA corrections cas in Eq.~(\ref{sugraexpan}).
Expanding the field $\phi$  in a Taylor series around the critical point,  the potential takes the form
\begin{equation}
        V=   \Lambda\left[ \left(1-\frac{\psi^2}{M^2}\right)^2+ \frac{2 \phi^2 \psi^2}{M^4 } +a_0+a_1 (\phi- \phi_c)+ a_2 (\phi- \phi_c)^2+a_3 (\phi- \phi_c)^3+a_4 (\phi- \phi_c)^4\right],
\end{equation}
where {$a_0=0.00707364$, $a_1=4.331\times 10^{-7}\,  M_P^{-1}$, $a_2=-0.0088\,  M_P^{-2}$, $a_3=0.0242\,  M_P^{-3}$ and $a_4=0.118\,  M_P^{-4}$. These values correspond to the parameters $M=0.05\, M_P$, $b_1=3.506\times10^{-4}M_P$ and $b_2=-3.5\times 10^{-3}$,}
as in the set 1.
    
Evaluating  the potential derivatives, the equations of motion read as
\begin{equation}
      3H\dot\phi =- \Lambda a_1\left( 1+\frac{4\psi^2 \phi}{M^4 a_1}\right)
    \label{eoms01}
\end{equation}
\begin{equation}
  3H\dot \psi=-\frac{4\psi\Lambda}{M^2}\left( \frac{\phi^2-M^2}{M^2}+ \frac{\psi^2}{M^2}\right).
    \label{eoms02}
\end{equation}
In order to solve analytically these equations, we follow 
the standard procedure to divide the inflationary period into three phases~\cite{Kodama:2011vs,Clesse:2015wea,Clesse:2013jra}. 
 In the so-called  phase 0, we neglect the second term of rhs of Eq.(\ref{eoms01}) and the first term of rhs of Eq.(\ref{eoms02}). 
 Consequently in the phase 1 the first term of rhs of Eq.(\ref{eoms01}) as well as the first term  of Eq.(\ref{eoms02}) are dominant. 
 Finally in the phase 2 the dominant terms are  the second term of rhs of Eq.(\ref{eoms01}) and the first term of rhs of Eq.(\ref{eoms02}).
  We omit the terms proportional to $a_2$, $a_3$ and $a_4$ in Eq.~(\ref{eoms01}) because in phases 0  and 1 $\phi$ varies quite  slowly with respect to   $\phi_c$. Moreover, in phase 2 this term does not contribute.
   As it is shown in Refs.~\cite{Kodama:2011vs,Clesse:2013jra}, the duration of the phase 0 is small.
Therefore, we consider only phases 1 and 2.

For convenience we introduce two new variable $\xi,\chi$, which are related to $\phi,\psi$ as:
\begin{equation}
\begin{split}
\phi=& \phi_c e^{\xi}\\
\psi=&\psi_0e^{\chi}.
\end{split}
\end{equation}
During the waterfall, as long as the slow-roll approximation is valid ($| \chi|  \ll 1$), we use 
 the approximation $\phi\simeq \phi_c(1+\xi)$. Moreover, in Eq.~(\ref{fcrit}), we assume that the value of the critical point is  $\phi_c=M$.

The equations of motion  (\ref{eoms01}) and (\ref{eoms02}) during the    two phases, phase 1 and phase 2 get the form
\begin{itemize}
\item Phase 1:  
\begin{equation}
\begin{split}
3H \dot \xi =& -\Lambda \frac{a_1}{M}    \, , \\
3H \dot \chi=& -\frac{4 \Lambda }{M^2}(2 \xi ) \, ,
\end{split}
\label{eomphase1}
\end{equation}
\item Phase 2:  
\begin{equation}
\begin{split}
3H \dot \xi =& -\Lambda\left( \frac{4 \psi^2}{M^4 }\right)  \, , \\
3H \dot \chi=& -\frac{4 \Lambda }{M^2}\left(2 \xi \right) \, .
\end{split}
\label{eomphase2}
\end{equation}

\end{itemize}

Thus,  the field trajectory through the   phase 1,  is governed  by Eqs.(\ref{eomphase1}):
$\frac{d \xi}{d \chi} \frac{4} {M^2} (2\xi)=  \frac{a_1}{M}$, so  $2 \xi d \xi = \frac{M a_1}{4 }d \chi$ with solution  
\begin{equation}
 \xi^2 =\frac{M a_1}{4 } \chi.
\label{ksi2} 
\end{equation}
The number of e-folds during phase 1 of inflation can be evaluated from Eqs. (\ref{eomphase1}). Hence,  one gets 
\begin{equation}
N_{1}(\xi)= -\frac{\xi M }{ a_1 M_P^2} \, .
\end{equation}
 \noindent
Assuming that there is exact match between phase 1 and 2, 
we get  for the field $\chi$
\begin{equation}
    \chi_2 = \ln \Big(   \frac { M^{3/2}\sqrt{a_1}}{2 \psi_0} \Big).
\end{equation}
Hence from the second equation of (\ref{eomphase1}), we derive:
\begin{equation}
N_1 =\frac{\chi_2^{1/2} M^{3/2}}{2M_{p}^2\sqrt{a_1}},
\end{equation}
which gives the total number of e-folds during phase 1. This number of e-folds is valid, if $\chi_2>m^2a_1 /8$~\cite{Kodama:2011vs}.

As for the phase 2 the solution of (\ref{eomphase2}) is
\begin{equation}
\xi^2=\xi_2^2 +\frac{M a_1}{4 }(e^{2(\chi-\chi_2)}-1)
\end{equation}
and  from Eq.~(\ref{ksi2})
\begin{equation}
\xi_2 = -\frac{ \sqrt{a_1 \chi_2 M}}{2}  \, ,
\end{equation}
where we match the phase 1 with phase 2. The total number of efolds during phase 2 
is given to a good approximation by~\cite{Kodama:2011vs,Clesse:2015wea,Clesse:2013jra}:
\begin{equation}
N_2 = \frac{M  {\phi_c}^{1/2}}{4 \, M_P^2\, a_1^{1/2}\,  x_2^{1/2}} \, .
\end{equation}

For the evaluation of the scalar power spectrum we have  in the $\delta N$ formalism~\cite{Clesse:2013jra,Sasaki:1995aw,Fujita:2014tja}:
\begin {equation}
P_R= \frac{H^2}{4 \pi^2} \Big(   N^2_{,\psi} +N^2_{,\phi}  \Big),
\end{equation}
 \noindent
 where $N_{,\phi}=N_{1,\phi}+N_{2,\phi}$ and $N_{,\psi}=N_{1,\psi}+N_{2,\psi}$ are the partial derivatives with respect to  fields $\psi$ and $\phi$. The subscripts $1,2$ denote the phases 1 and 2 respectively. It has been shown \cite{Clesse:2013jra} that the partial derivatives in phase 2 do not have a significant contribution. Hence we neglect them and we evaluate those in phase 1. For the derivatives we get:
 \begin{equation}
    N_{,\phi}=-\frac{ 1}{ a_1 M_P^2}, \quad  {N_{,\psi}=- \frac{M^{2}}{8 \, M_P^2 \, \xi_2 \, \psi_k}.}   
 \end{equation}
So the  power spectrum in the slow-roll approximation  is 
 \begin{equation}
{ P_R = \frac{1}{4 \pi^2}\frac{\Lambda}{3M_P^2}\left(\frac{  1}{ a_1^2 M_P^4} 
 +\frac{M^4}{64M_P^4 \, \xi_2^2\,  \psi_k^2}\right)    
     \approx \frac{1}{4 \, \pi^2} \frac{\Lambda}{3\, M_P^2} \frac{M^4}{64 \, M_P^4 \, \psi_k^2 \, \xi_2^2}\, ,}
 \end{equation}
where, as it has pointed out in~\cite{Clesse:2013jra}, 
 in  a good approximation we have also   neglected the term $ N_{,\phi}$\cite{Clesse:2013jra}. 
 With $\psi_k$ we denote the following quantity 
\begin{equation}
\psi_k =\psi_0 \, e^{\chi_k},
\end{equation}
where 
\begin{equation}
\chi_k=\frac{4a_1M_P^4}{M^3}\, (N_1+N_2-N)^2.
\end{equation}
Eventually,  the analytical expression for the power spectrum we use  is  
\begin{equation}
P_R= \frac{\Lambda M^3   }{192 \, \pi^2a_1 \, M_P^6 \, x_2 \, \psi_k^2}  \, .
\label{analytical}
\end{equation}

\subsection{Numerical solution  beyond the slow-roll}

In order to check our   analytical calculation  for the scalar power spectrum, 
 we also evaluate it  numerically.
 Thus, in this section  we present this numerical evaluation  for the perturbation of the fields and  the power spectrum.  
 This way we will check  that the previous analytical calculation can  be regarded as a good approximation.

The perturbed metric is given by
\begin{equation}
ds^2=a^2 \left[ -(1+2 \Phi)d \tau^2+\left( \left(1-2 \Psi\right)\delta_{ij}+\frac{1}{2}h_{ij}\right) dx^i dx^j \right]
\label{eq.metr}
\end{equation}
where $h_{ij}$ are the  tensor perturbations. 
With $\Phi$ and $\Psi$ we denote the Bardeen potentials which are equal in the conformal-Newtonian gauge.

For evaluating the scalar perturbations, which are denoted as $\varphi_i+\delta\varphi_i$, we assume the 
linear equations
\begin{equation}
\delta  \varphi_i ^{\prime \prime}+(3-\epsilon) \delta  \varphi_i ^\prime+\sum \frac{1}{H^2}\frac{\partial ^2 V}{\partial \varphi_i \partial \varphi_j}\delta \varphi_i+\frac{k^2}{a^2H^2}\delta \varphi_i=4 \Phi'\varphi_i'- \frac{2\Phi}{H^2}\frac{\partial V}{\partial \varphi_i}\, ,
\label{eq.pert}
\end{equation}
where the subscript $i$ refers to the fields $\phi,\psi$ and the Bardeen potential $\Phi$ is given by the solution of equation
\begin{equation}
\Phi^ {\prime \prime}+(7-\epsilon)  \Phi^ \prime+ \left(    2 \frac{V}{H^2} + \frac{k^2}{a^2H^2} \right)\Phi=-\frac{V_{,\varphi}}{H^2}.
\label{eq.Bardeen}
\end{equation}
With $k$ we denote the comoving wavenumber and both Eqs.~(\ref{eq.pert}) 
and (\ref{eq.Bardeen}) are expressed  in terms of  the number of e-folds. Primes denote derivatives 
in e-fold time. Integrating  this equation, we evaluate the scalar power spectrum 
using  the  expression
\begin{equation}
\label{eq1.5}
P_R=\frac{k^3}{2 \pi^2} \left|R_k \right|^2,
\end{equation}
where $R_k$ is the comoving curvature perturbation:
\beq
R_k=\Phi +\frac{\delta \varphi }{\varphi'}.
\eeq
The initial conditions of these equations as well as the numerical treatment of them are found in Ref. \cite{Ringeval:2007am}.
  \begin{figure}[h!]
\centering
\includegraphics[width=100mm]{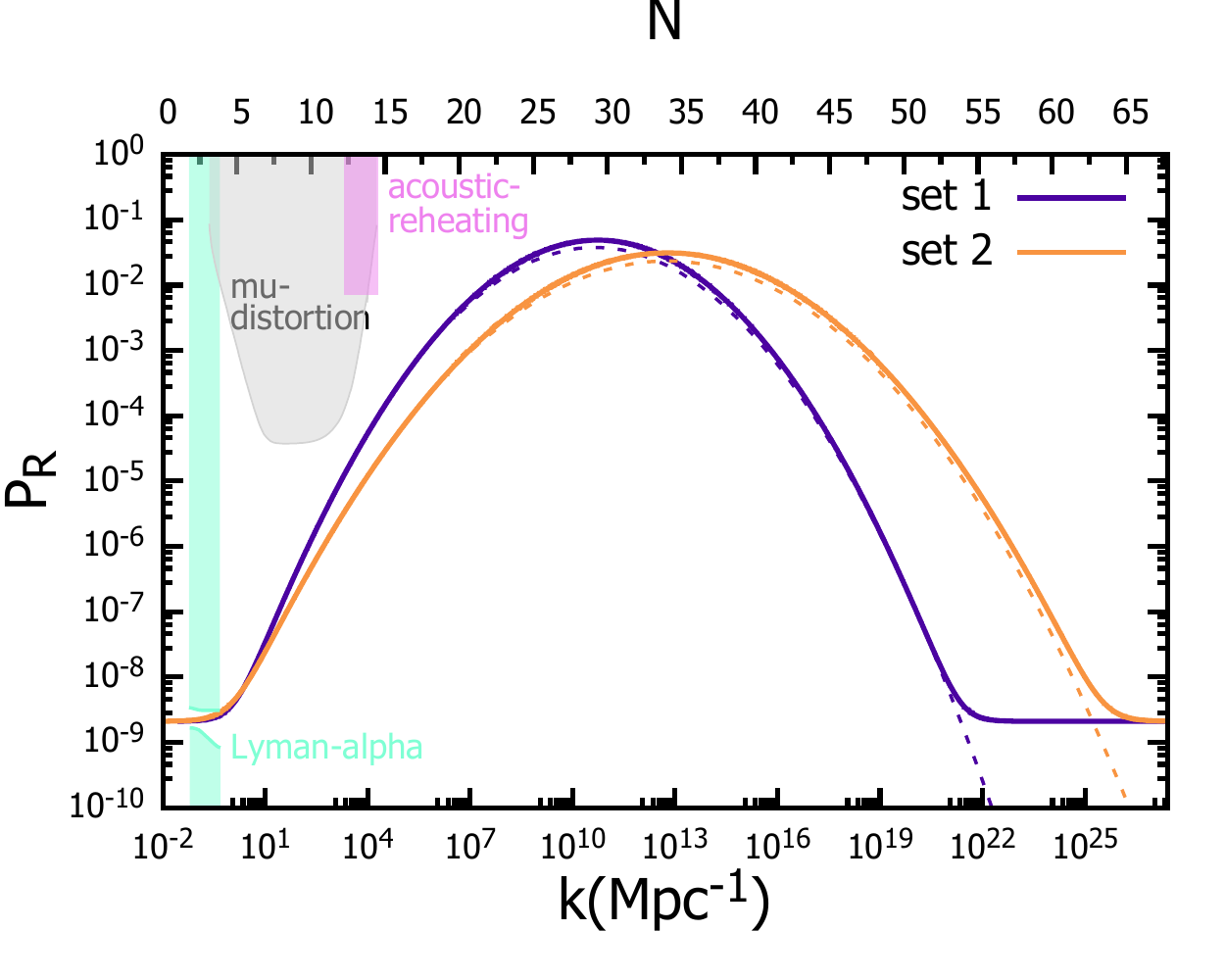}
\caption{The power specta in the slow-roll approximation (solid lines)  and using the  exact  equation~(\ref{eq1.5}) of perturbations (dashed lines). 
Purple line corresponds to the set 1 and orange  line to set 2, given in Table~\ref{tabunsr}. Details are on main text. }
\label{f6}
\end{figure}

 In Fig.~\ref{f6} we present  the exact (numerical)  and analytical (in the slow-roll approximation)  power spectrum for the potential (\ref{eq.potsugra}). 
{In this plot we also depict for comparison reasons the current bounds from  the Lyman-alpha forest~\cite{Lymanalpha}, the mu-distortion~\cite{Fixsen:1996nj} and the acoustic-reheating bound~\cite{Nakama:2014vla,Jeong:2014gna}.}
The exact results are plotted as dashed  lines  and the slow-roll analytical results as solid lines. 
{ For the analytical expression we use the Eq.~(\ref{analytical}) and for the numerical expression we use the Eq.~(\ref{eq1.5}). For the numerical evaluation we follow the analysis described in Refs.~\cite{Clesse:2015wea,Clesse:2013jra}. Specifically, we solve numerically the background equations (\ref{eq:bckgeq}), the fields' perturbation (\ref{eq.pert}) and the Bardeen potential (\ref{eq.Bardeen}) simultaneously until the end of inflation (the parameter $\varepsilon$ reaches value 1) starting from  the critical point as initial condition. We obtain at around 27 efold, as it is shown in  Fig.\ref{f5}. We integrate again the system of the different equations, (\ref{eq:bckgeq}), \ref{eq.pert}) and   (\ref{eq.Bardeen}), until $k=0.05 Mpc^{-1}$ and we find the initial conditions and we obtain the rest number of efold. Finally, we integrate again the system from the new initial conditions until the end of inflation. }
We  notice that the analytical and the numerical results are very similar, 
as it is expected~\cite{Clesse:2015wea}. 
Therefore for convenience we shall 
 use the analytical slow-roll solution hereinafter.
%


\section{Gravitational Waves Production}

	In the previous sections,
	 we have presented a mechanism in the context of a  SUGRA based hybrid model, 
	  that can produce  a  significant enhancement in the scalar power spectrum.  
	 The amount of GWs is evaluated by the second-order (tensor)  perturbations, 
	 which appear as $h_{ij}$   in Eq.~(\ref{eq.metr}). However, the tensor second-order perturbations can 
	 be related to the scalar first-order perturbations, and hence to the scalar power spectrum~\cite{Acquaviva:2002ud,Mollerach:2003nq,Ananda:2006af,Baumann:2007zm,Espinosa:2018eve,Kohri:2018awv,Matarrese:1997ay}.  
	 In this section we show that the enhancement of the power spectrum can be interpreted as a source 
	 of GWs created during the radiation dominance era.
 \begin{figure}[h!]
\centering
\includegraphics[width=100mm]{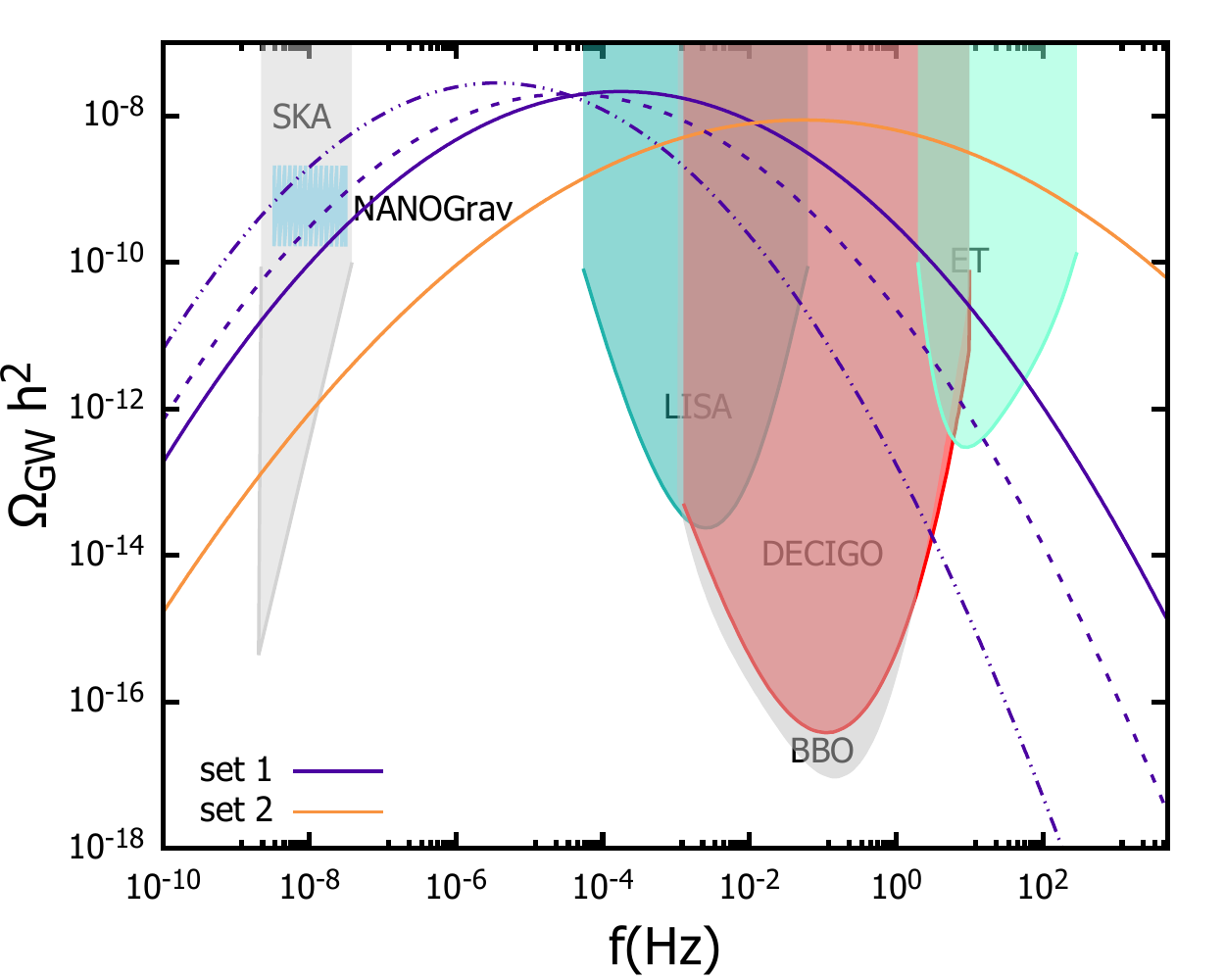}
\caption{ The energy density of gravitational waves for the analytical expression Eq.~(\ref{analytical}).  Purple  curves  correspond
 to set 1 and the orange to set 2, as given in Table~\ref{tabunsr}.  
 Detailed description of the various curve shapes can be found in the main text.}
\label{f7}
\end{figure}

The present-day energy density of the GWs is~\cite{Maggiore:1999vm}
\begin{equation}
{\Omega_{GW}(k)}=\frac{1}{24}\left( \frac{k}{aH}\right)^2 \overline{P_h(\tau,k)} \,,
\end{equation}
where $P_h$ is the tensor perturbation and the over-line denotes the average over the time. In terms of scalar power spectrum this expression reads as \cite{Espinosa:2018eve}:
\begin{equation}
{\Omega_{GW}(k)}=\frac{c_g\, \Omega_r}{36} \int^{\frac{1}{\sqrt{3}}}_{0}\mathrm{d} d
\int ^{\infty}_{\frac{1}{\sqrt{3}}}\mathrm{d} s \left[  \frac{(s^2-1/3)(d^2-1/3)}{s^2+d^2}\right]^2 \, 
P_{R}(kx)P_{R}(ky)(I_c^2+I_s^2).
\label{eq4.1}
\end{equation}
The radiation density $\Omega_r$ gets its measured present day  value $\Omega_r = 5.4  \times 10^{-5}$  and $c_g=0.4$ 
in  the case of Standard Model (SM)  spectrum, while $c_g=0.3$ 
in the Minimal Supersymmetric SM (MSSM). 
The variables $x$ and $y$ are:
\begin{equation}
x= \frac{\sqrt{3}}{2}(s+d), \quad  y=\frac{\sqrt{3}}{2}(s-d).
\label{eq4.2}
\end{equation}
Finally, the functions $I_c$ and $I_s$ are given by the equations
\begin{gather}
I_c=-36 \, \pi \, \frac{(s^2+d^2-2)^2}{(s^2-d^2)^3} \, \Theta(s-1)\\
I_s=-36 \, \frac{(s^2+d^2-2)^2}{(s^2-d^2)^2} \Bigg[ \frac{(s^2+d^2-2)}{(s^2-d^2)} \, \ln \left| \frac{d^2-1}{s^2-1} \right| +2 \Bigg] \, .
\label{eq4.3}
\end{gather}
Using that  $1 \, \mathrm{Mpc}^{-1}= 0.97154\times 10^{-14} \, \mathrm{s}^{-1}$ and $k=2\pi f$, 
we can evaluate the energy density of the GWs as a function of the frequency. 

In Fig.~\ref{f7} we plot the energy density of GWs using the analytical expression in 
Eq.~(\ref{analytical}).  Purple  curves  correspond
 to set 1 and the orange to set 2, as given in Table~\ref{tabunsr}.  
 Moreover, the solid (dashed, dashed-double dot) purple  line correspond  to $a_1=4.33$ ($a_1=5$, $a_1=6$),
 in units $10^{-7}/M_P$.
 The parameter $a_1$ is defined  in (\ref{sugraexpan}). 
As can be seen from the figure,  
the predicted GW spectra for these  parameter choices, in the context of our hybrid model, 
lie well within the detection range of 
 the future 
GW  experiments, like LISA, DESIGO, BBO, SKA and ET~\cite{Audley:2017drz,Sato:2017dkf,Sathyaprakash:2009xs,Zhao:2013bba,Yagi:2011wg}.
Interestingly enough, we notice  that    the recently reported    NANOGrav \cite{Arzoumanian:2018saf,Aggarwal:2018mgp,Arzoumanian:2020vkk}    signal
 of can be interpreted in the context of this model  (purple lines). 
  In this figure we display  the NANOGrav 12.5 yrs  region.
  %

\section{Primordial Black Holes Abundance}
The significant enhancement of scalar power spectrum not only can explain the energy density of GWs, 
as it is was discussed in the previous section, but also  the production of the PBHs.
 A main result  of Ref.~\cite{DeLuca:2020agl} is that the GWs spectrum , 
which is compatible to the NANOGrav region, can be related  to a particular prediction for the PBH abundance. 
In this section we evaluate the mass of PBHs and their fractional abundances.
As usual it is assumed   that the PBH are formed    in the radiation 
dominated epoch, as the GWs.

The fractional abundance of PBHs, $\Omega_ {PBH} / \Omega_ {DM}$, can be evaluated as a 
function of the PBH mass using that
\begin{equation}
\label{44}
\frac{\Omega_ {PBH}}{\Omega_ {DM}}(M_{\mathrm{PBH}})= 
\frac{\beta(M_{\mathrm{PBH}})}{8 \times 10^{-16}} \left(\frac{\gamma}{0.2}\right)^{3/2} 
       \left(\frac{g_*(T_f)}{106.75}\right)^{-1/4}\left(\frac{M_{\mathrm{PBH}} \,  }{10^{-18 }\; \mathrm{ grams} }\right)^{-1/2}\, , 
\end{equation}
where we assume that the DM abundance is $\Omega_{DM} \simeq 0.26$. With $\gamma$ we denote a factor which depends on the gravitation collapse and we choose $\gamma =0.2$~\cite{Carr:1975qj}. With $\beta$ we denote the mass fraction of Universe collapsing in PBH mass. The $T_f$   denotes  the temperature  of PBH formation and the    $g_*(T_f)$  are the effective degrees of freedom during this formation. In order to evaluate the  abundance of PBHs, we integrate the expression in Eq.~(\ref{44}) as
\beq
f_ {PBH} = \int \frac{d  M_{\mathrm{PBH}}}{M_{\mathrm{PBH}}} \, \frac{\Omega_ {PBH} }{\Omega_{DM}}\,  .
\label{fpbh}
\eeq

The mass of PBHs, which are created after the inflation when the scales reenter the horizon is related to the mass inside the Hubble horizon. Specifically, the mass of PBHs is 
\beq
M_{\mathrm{PBH}}=\gamma \frac{4\, \pi \,  \rho}{3}  H^{-3} \,  ,
\eeq
where $\rho$ is the energy density of Universe during collapse.
If we consider that the PBHs are formed during the radiation epoch, their mass is~\cite{Ballesteros:2017fsr}
\begin{equation}
M_{\mathrm{PBH}}=\gamma \frac{4\, \pi \, \rho}{3}\,  H_{m-r}^{-3} \left( \frac{g(T_f)}{g(T_{m-r})}\right)^{1/2}\left( \frac{g_s(T_f)}{g_s(T_{m-r})}\right)^{-2/3}\left( \frac{k}{k_{m-r}}\right)^{-2}  \, ,
\end{equation}
where the subscript ${m-r}$ refers to the time of equality of matter and radiation domination and $g_s$ refers to the entropy density. 
The equation above arises from the entropy  conservation  $d(g_s(T)T^3a^3)/dt=0$ between the epoch of the reentry 
of the comoving wavenumbers and the epoch of radiation-matter equality.
Thus we can express   the mass of PBHs as a function of the comoving wavenumber $k$
\begin{equation}
\label{43}
M_{\mathrm{PBH}}(k)=10^{18} \left( \frac{\gamma}{0.2} \right) \left(\frac{g_*(T_f)}{106.75}\right)^{-1/6} \left(\frac{k}{7 \times 10^{13} \, \mathrm{Mpc}^{-1}  }\right)^{-2}  \mathrm{ in~grams} \, , 
\end{equation}
where we use the approximation    $g(T)=g_s(T)$~\cite{Ballesteros:2017fsr}. 
Assuming  that the spectrum of the our model 
 is like  the SM,  we  can use  ${g_*(T)=106.75}$. 
 On the other hand assuming a spectrum like the MSSM,
  we get  $g_*(T)=228.75$. 
  Thus the PBH fractional abundance in the SM is 1.13 times larger than in the MSSM.
  This relative factor to a good approximation can be ignored. 
  

\begin{figure}[h!]
\centering
\includegraphics[width=100mm]{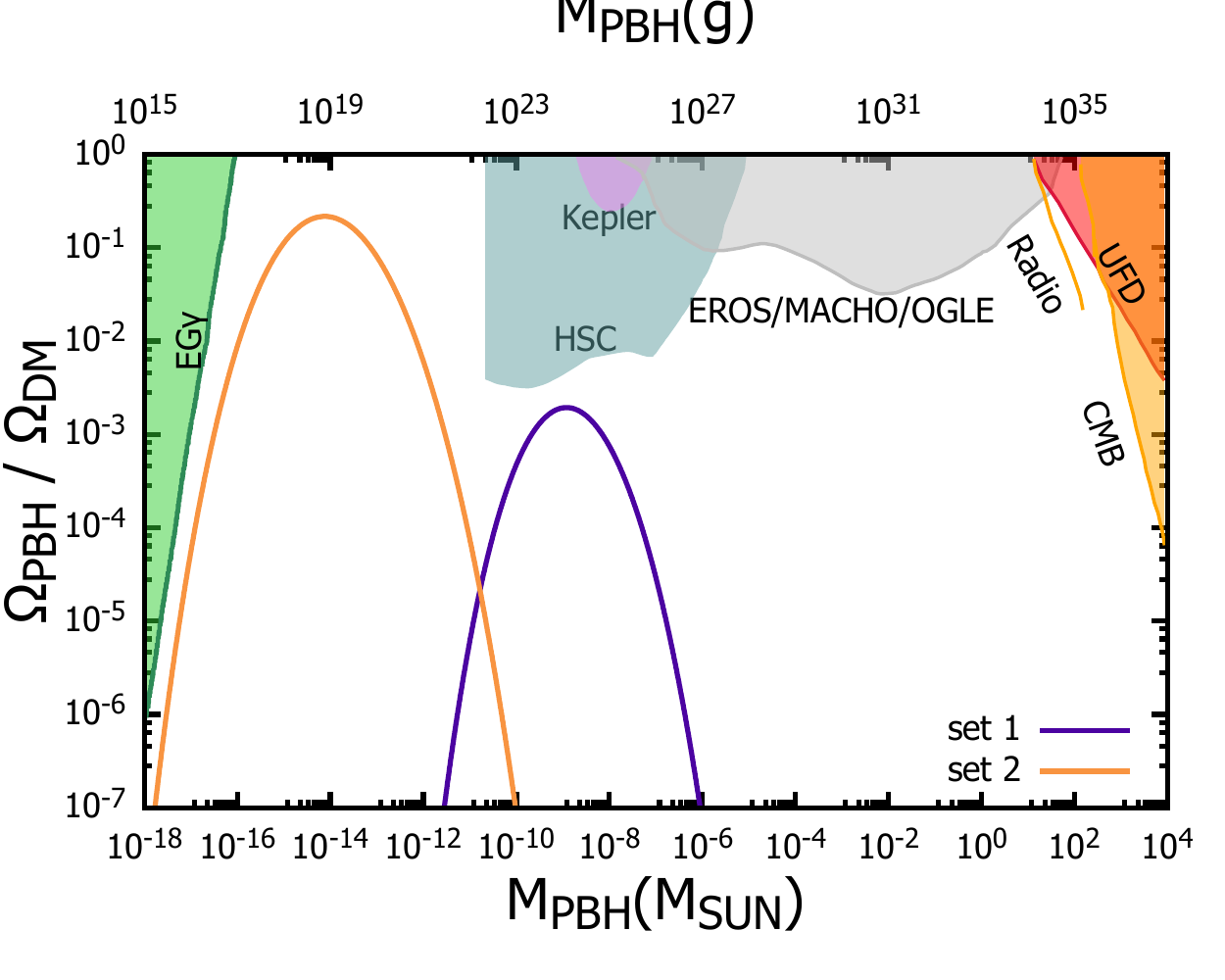}
\caption{ The fractional abundance of PBHs as a function of mass.  
As before, the purple   line corresponds to set 1 and orange line to set 2 given in Table \ref{tabunsr}.}
\label{f8}
\end{figure}

The mass fraction $\beta$ is evaluated  using  the Press-Schechter approach. 
In this approach,   the  mass fraction $\beta$ is calculated assuming that  the overdensity $\delta$ follows a gaussian   probability, 
with a  threshold of  collapse $\delta_c$.
 So the mass fraction is given from the  integral
\begin{equation}
\label{42}
\beta(M_{\mathrm{PBH}})= \frac{1}{\sqrt{2 \pi \sigma ^2 (M_{\mathrm{PBH}})}} \int^{\infty}_{\delta_c} d\delta \,  \exp \left(  -\frac{\delta ^2}{2 \sigma^2(M_{\mathrm{PBH}}) } \right) \, , 
\end{equation}
where $\sigma$ is the variance of curvature perturbation,   related to the comoving wavenumber as
\begin{equation}
\label{40}
\sigma^2 \left( M_{\mathrm{PBH}}(k)  \right)= \frac{16}{81}  \int \frac{dk' }{k'} \left(\frac{k'}{k}\right)^4 P_{R}(k') \tilde{W}\left(\frac{k'}{k}\right).
\end{equation}
{ where  $\tilde{W}(x)$ is a window function. We consider the Gaussian distribution for this function: $ \tilde{W}(x)=e^{-x^2/2} $. } For   $\delta_c$, following recent studies~\cite{Harada:2013epa,Musco:2008hv,Musco:2004ak,Musco:2012au,Musco:2018rwt,Escriva:2019phb,Escriva:2020tak,Musco:2020jjb},
 we consider values in the range  between 0.4 and 0.6.

In Fig.~\ref{f8} we plot the fractional abundance of  PBHs from Eq.~(\ref{44}). 
We use the analytical expression for the scalar power spectrum from Eq.~(\ref{analytical}).
 In this plot  we use the value of the $\delta_c= 0.50$  for the  set 1 (purple curve) and $\delta_c=0.43$ for the set 2 (orange curve).
  In both cases,   a significant fractional abundance of PBHs is obtained. 
To form an idea of the allowed parameter space, 
in the Fig.~\ref{f8} we also display
the disallowed regions due to  various 
observational groups and studies~\cite{Carr:2009jm,Inoue:2017csr,Montero-Camacho:2019jte,Katz:2018zrn,Poulin:2017bwe,Capela:2013yf,Niikura:2017zjd,Wyrzykowski:2011tr,Griest:2013esa,Tisserand:2006zx,Ali-Haimoud:2016mbv,Gaggero:2016dpq}.

Specifically, we   notice a tension between  the  NANOGrav data for the GW spectrum 
and the SUBARU HSC  experiment exclusion limit for the PBH fraction.  
As  a result of this, for the  set 1 (purple curve) the calculated abundance of PBHs from Eq.~(\ref{fpbh}) is ${f_{PBH} \simeq 0.01}$,
while for  set 2 (orange line) we get  $f_{PBH} \simeq 1$. That is,  in former case PBH can account just for the ${1\%}$ of the DM of the Universe,
while in the latter can be almost $100\%$. 
In order to understand this one has to compare  Fig.~\ref{f8} with  Fig.~\ref{f7}. 
In particular, in Fig.~\ref{f7} the frequency $f$ is proportional to comoving wavenumber $k$, but the $M_{\mathrm{PBH}}$ in  Fig.~\ref{f8} is
proportional to $k^{-2}$. Thus, to fulfil the NANOGrav region result to move the GW spectrum to the right close to the 
disallowed regions. 
In any case, the hybrid model under consideration  can be regarded as proper candidate to explain the whole DM in the Universe. 
Finally, hybrid models can also explain the production of PBHs in a wide range of masses. 

\section{Fine-tuning estimation  and Conclusions}

A drawback of the models, where  an enhancement of scalar power spectrum is produced due to 
a modification of the potential, is the high level of required fine-tuning of
the parameters involved in this modification. 
In particular,  models,  where an inflection point in effective scalar potential is developed, demand  a significant amount of fine -tuning~\cite{Stamou:2021qdk,Hertzberg:2017dkh}. 
In this section we perform an analogous study for the hybrid model under consideration and we
show that the corresponding amount of the fine-tuning is much smaller,  at least by 3 or 4 orders of magnitudes.   

As discussed in Section~\ref{sect:model}, the basis of our hybrid model is the original hybrid model introduced in \cite{Linde:1993cn}, 
with the addition of  extra SUGRA terms, that correct the value of the spectral index $n_s$.
Hence,  we have had two 
extra parameters, $b_1$ and  $b_2$. Generally speaking, using  the parameter $b_2$ we decreased the value of the $n_s$ 
to be in accordance with the observable constraints, while the other  parameter $b_1$ is used
  for explaining the enhancement in power spectrum and then the production of GWs and PBHs.

In order to have a measure  of the required fine-tuning in this model, 
we calculate  the quantity $\Delta_{p}$, as discussed in~\cite{Barbieri:1987fn},  for the parameter $b_1$. 
$\Delta_{p}$  is the maximum value of the  logarithmic derivative of the peak value of the power spectrum, with respect to $b_1$
\begin{equation}
\Delta_p=\mathrm{Max} \left| \frac{\partial \ln(P_R^{PEAK}) }{\partial \ln(b_1) }   \right|.
\label{deltaeenz}
\end{equation}
The larger the $\Delta_{p}$, the larger the amount of the required fine-tuning. 
Evaluating  numerically $\Delta_{p}$ from  Eq.~(\ref{deltaeenz}),  for the function $P_R^{(peak)} (b_1)$ 
we find that $\Delta_{p}   \sim 100$, if we demand a peak of power spectrum at around $5 \times 10^{-2}$. 
Thus,  we conclude that  the amount of the fine -tuning is  significantly smaller, by almost four 
orders of magnitude   in the case of this hybrid model, compared 
to the single field inflation models, where an inflection point  is the source for producing PBHs~\cite{Stamou:2021qdk}. 
 Finally, we notice here that the fine-tuning for an acceptable values for the $n_s$ is mainly  based  on the parameter $b_2$ and this issue. For analogous equation of Eq.~(\ref{deltaeenz}) we obtain a quantity at $ \Delta_p \sim10$  in the range near to $b_2$.

Concluding, 
in this work we presented a two-field inflationary model based on the original simple  hybrid model, in order to explain the generation of both GWs and PBHs.  
The effective scalar potentials derived by hybrid models have the advantage that they do not require a high level of fine-tuning of 
the parameters in order to describe an amplification in scalar power spectrum. As the issue of fine-tuning is regarded as a main problematic feature in 
many proposed models, studying hybrid models should be a plausible scenario in order to describe enhancement in scalar power spectrum. 
A disadvantage of these models is that they cannot predict acceptable values for the spectral index $n_s$. 
For this reason we introduced specific SUGRA-type  corrections in the inflaton field, where we evaluated the prediction for the cosmological 
 constraints. In our proposed model we have achieved acceptable values for both spectral index and tensor-to-scalar ratio.

Hybrid models can explain the production of GWs and PBHs due to  the amplification of the scalar power spectrum. 
This enhancement occurs because of  the mild waterfall of one of the two fields. We presented an analytical procedure for 
evaluating the scalar power spectrum of our proposed model based on slow-roll approximation. For comparison,
 we evaluated numerically  the exact equations of perturbations by using the exact proposed potential. 
 We conclude that the results for analytical and numerical procedure are quite similar,  since they differ up to $10\%$.

Having calculated the scalar power spectrum  we evaluated the amount of the produced 
GWs in the radiation 
dominated epoch. 
To facilitate the phenomenological analysis of our model we have chose two representative 
sets of  the two parameters  for the extra terms in the  K\"ahler potential.
The first   set corresponds to a model that predicts GW spectra that lie within the NANOGrav  detection region.
Unfortunately the associated factional abundance of  PBHs is restricted by the SUBARU HSC exclusion region. 
As a result in this case only 1\% of DM can be interpreted by  PBHs.  On the other hand, in the second set 
we relax the NANOGrav constrain. This way we are able to increase the  factional abundance of  PBHs up to  100\%.

\appendix
\section{Appendix}
\label{sect:app}

In order to evaluate the SUGRA corrections in the scalar potential, we consider the K\"ahler potential:
\begin{equation}
{     K= S \bar{S} +b_1(S +\bar{S}) +b_2(S +\bar{S})^2  +\Psi_1{\bar{\Psi}_1}+\Psi_2{\bar{\Psi}_2}}\,
    \label{eqkahler1}
\end{equation}  
and we assume that the superpotential is given from \begin{equation}
W= \kappa \, S  (  \Psi_1{\Psi}_2  - \frac{M^2}{2})\, 
\label{spotential2}
\end{equation} 
We calculate the $F$-term of scalar potential from Eq.~(\ref{eqvgeneral}) and we find
\begin{equation}
{\begin{split}
  V_F^{\text{SUGRA}} & ={ \frac{\kappa^2}{4 {M_P}^4}  }      \exp \Bigg[ \frac{1}{   M_P^2} \left(\frac{b_1 \phi }{\sqrt{b_2+\frac{1}{2}}}+\frac{{b_2} \phi ^2}{b_2+\frac{1}{2}}+\frac{\phi ^2}{4 b_2+2}+{\psi ^2}\right) \Bigg]
\\ &  \times     \Bigg[
   \frac{\left(M^2-\psi ^2\right)^2 \left({b_1} \sqrt{4 {b_2}+2} \phi +(4 {b_2}+2) {M_P}^2+(4 {b_2}+1) \phi ^2\right)^2}{4 (2 {b_2}+1)^3}  \\
   & +\frac{\phi ^2 \psi ^2
   \left(-M^2+2 {M_P}^2+\psi ^2\right)^2 }{4 {b_2}+2}-\frac{3 {M_P}^2 \phi ^2 \left(M^2-\psi ^2\right)^2}{4 b_2+2} \Bigg]
   \end{split}}
\label{eq.ful.sugra}
\end{equation}
where we have fix the non-canonical kinetic terms by the definition of the chiral fields
\begin{equation}
S=\frac{\phi}{\sqrt{2+4b_2}}, \quad |\Psi_1|=|{\Psi}_2|={\psi/\sqrt{2}}.
\label{eq.hy.fix.nncan}
\end{equation}
Moreover, we derive the scalar potential   using the general form
\begin{equation}
\begin{split}
   V_F^{\text{SUGRA}}&= \left(1+\frac{K}{M_P^2}+\frac{K^2}{2M_P^4}+\frac{K^3}{6M_P^6}\right)\\
 & \times  \left[\left( {K^{-1}}\right)^i_{\bar{j}} \left(W^{\bar{j}}+\frac{WK^{\bar{j}}}{{M_P}^2}\right) \left(\bar{W}_i+\frac{\bar{W}K_{i}}{{M_P}^2}\right)- \frac{3|W|^2}{{M_P}^2} \right] 
   \label{eq:ap2}
   \end{split}
\end{equation}
and we have 
\begin{equation}
V_F^{\text{SUGRA}}=\frac{\kappa^2(M^4-2M^2\psi^2+2\psi^2\phi^2+\psi^4)}{4+8b_2}+\frac{\mathcal{A}_1}{M_P^2}+ \frac{\mathcal{A}_2}{M_P^4}+\frac{\mathcal{A}_3}{M_P^6}+\mathcal{O}\left(\frac{1}{M_P^8}\right)
\label{eq.expn}
\end{equation}
In the limit $M_P\rightarrow \infty$ we derive the SUSY scalar potential. 
The SUGRA corrections $\mathcal{A}_1$, $\mathcal{A}_2$ and $\mathcal{A}_3$ are calculated as
\begin{equation}
\begin{split}
\mathcal{A}_1 =&\frac{\kappa^2}{4 (2b_2+1)^{5/2}}\Big[2 \sqrt{2} {b_1} (2 {b_2}+1) \phi \left(M^4-2 m^2 \psi^2+\phi^2 \psi^2+\psi^4\right)+\\
+&2 \sqrt{2} {b_1} (2 {b_2}+1) \phi \left(M^4-2 M^2 \psi^2+\phi^2 \psi^2+\psi^4\right)+\\
+& {b_2} \Big(M^4 \left(3 \phi^2+2 \psi^2\right)+2 M^2 \psi^2 \left(-5 \phi^2-2
   \psi^2+2M_P^2\right)+\\+&\psi^2 \left(4 \phi^4+11 \phi^2 \psi^2-2M_P^2 \left(2 \phi^2+3 \psi^2\right)+2
   \psi^4\right)\Big)\Big]
\end{split}
   \label{eq:apA1}
\end{equation}

\begin{equation}   
\begin{split}
\mathcal{A}_2 =&\frac{\kappa^2}{32 (2 {b_2}+1)^{7/2}}\Big[ 4 {b_1}^2 (2 {b_2}+1)^{3/2} \phi^2 \left(7 M^4-14 M^2 \psi^2+4 \phi^2 \psi^2+7
   \psi^4\right)+\\
   +&8 \sqrt{2} {b_1} \phi \Big(\left(8 {b_2}^2+6 {b_2}+1\right) \phi^4 \psi^2+5
   \left(6 {b_2}^2+5 {b_2}+1\right) \phi^2 \psi^4+\\+&(2 {b_2}+1) M^4 \left((7
   {b_2}+1) \phi^2+2 (2 {b_2}+1) \psi^2\right)-\\-&2 (2 {b_2}+1) M^2 \psi^2
   \left((9 {b_2}+2) \phi^2+2 (2 {b_2}+1) \psi^2\right)+2 (2 {b_2}+1)^2
   \psi^6\Big)+\\
   +&\sqrt{2 {b_2}+1} \Big(M^4 \big[\left(64 {b_2}^2+20 {b_2}+1\right)
   \phi^4+\\+&4 \left(16 {b_2}^2+10 {b_2}+1\right) \phi^2 \psi^2+4 (2 {b_2}+1)^2
   \psi^4\big]-\\-&2 M^2 \psi^2 \big[\left(96 {b_2}^2+44 {b_2}+5\right) \phi^4+12
   \left(8 {b_2}^2+6 {b_2}+1\right) \phi^2 \psi^2-\\-&8 (2 {b_2}+1)^2
   {M_P}^4+4 (2 {b_2}+1)^2 \psi^4\big]+\\+&\psi^2 \big[\left(192 {b_2}^2+116
   {b_2}+17\right) \phi^4 \psi^2+4 \left(40 {b_2}^2+34 {b_2}+7\right) \phi^2
   \psi^4-\\-&8 (2 {b_2}+1)^2 {M_P}^4 \left(2 \phi^2+3 \psi^2\right)+2 (4
   {b_2}+1)^2 \phi^6+4 (2 {b_2}+1)^2 \psi^6\big]\Big)\Big]
\end{split}
   \label{eq:apA2}
\end{equation}

\begin{equation}
\begin{split}
\mathcal{A}_3=&\frac{\kappa^2}{96 (2 {b_2}+1)^{9/2}}\Big[4 \sqrt{2} {b_1}^3 (2{b_2}+1)^2 \phi^3 \left(11 M^4-22 M^2 \psi^2+4 \phi^2 \psi^2+11
   \psi^4\right)+\\
   +&6 {b_1}^2 \sqrt{2 {b_2}+1} \phi^2 \Big(4 \left(8 {b_2}^2+6{b_2}+1\right) \phi^4 \psi^2+5 \left(32 {b_2}^2+26{b_2}+5\right) \phi^2 \psi^4+\\
   +&(2 {b_2}+1) M^4 \left((48 {b_2}+9) \phi^2+14 (2 {b_2}+1) \psi^2\right)-\\
   -&2 (2{b_2}+1) M^2 \psi^2 \big[(56 {b_2}+13) \phi^2+14 (2
   {b_2}+1) \psi^2\big)+14 (2 {b_2}+1)^2 \psi^6\Big)+\\
   +&6 \sqrt{2} {b_1} \phi \Big(2 M^4 \big(\left(56 {b_2}^3+50 {b_2}^2+13
   {b_2}+1\right) \phi^4+\\
   +&(16{b_2}+3) \phi^2 (2 {b_2} \psi+\psi)^2+2 (2   {b_2}+1)^3 \psi^4\big]-\\
   -&4 M^2 \psi^2 \left(\left(72 {b_2}^3+70{b_2}^2+21 {b_2}+2\right) \phi^4+5 (4 {b_2}+1) \phi^2 (2 {b_2} \psi+\psi)^2+2 (2 {b_2}+1)^3 \psi^4\right)+\\
   +&10 \left(24 {b_2}^3+26 {b_2}^2+9
  {b_2}+1\right) \phi^4 \psi^4+(2 {b_2}+1) \phi^6 (4 {b_2} \psi+\psi)^2+\\+&2 (28
{b_2}+9) \psi^6 (2 {b_2} \phi+\phi)^2+4 (2 {b_2}+1)^3 \psi^8\Big)+\\
+&\sqrt{2 {b_2}+1} \Big(M^4 \big[3 \left(160 {b_2}^3+144 {b_2}^2+38
 {b_2}+3\right) \phi^4 \psi^2+(4 {b_2}+1)^2 (17{b_2}+2) \phi^6+\\+
 &12 (5  {b_2}+1) \psi^4 (2 {b_2} \phi+\phi)^2+4 (2 {b_2}+1)^3 \psi^6\big]+\\
 +&\psi^2 \big[3\left(352 {b_2}^3+384 {b_2}^2+134 {b_2}+15\right) \phi^4 \psi^4-\\
 -&24 (2 {b_2}+1)^3 {M_P}^6 \left(2 \phi^2+3 \psi^2\right)+(4 {b_2}+1)^3 \phi^8+\\
 +&(41   {b_2}+14) \phi^6 (4 b_2 \psi+\psi)^2+4 (31 {b_2}+11) \psi^6 (2 {b_2}\phi+\phi)^2+4 (2 b_2+1)^3 \psi^8\big]-
   \\-&2 M^2 \psi^2 \big[-24 (2 {b_2}+1)^3
   {M_P}^6+(4 {b_2}+1)^2 (23b_2+5) \phi^6+\\+&21 (2 b_2+1) \phi^4 (4
   b_2\psi+\psi)^2+12 (7 b_2+2) \psi^4 (2 {b_2} \phi+\phi)^2+4 (2 b_2+1)^3
   \psi^6\big]\Big) \Big]
\end{split} 
\end{equation}

we derive that around the critical point $\phi_c=0.05M_P$ the numerical values of
 the corrections are:  $\mathcal{A}_1/{{M_P}^6\kappa^2} \approx 10^{-9}$,   
 $\mathcal{A}_2/{{M_P}^8\kappa^2} \approx 10^{-10}$ and  $\mathcal{A}_3/{{M_P}^{12}\kappa^2} \approx 10^{-12}$. 
 Moreover for these parameters, the spectral index  is   $n_s={0.96505}$. 
If we consider  corrections up to $\mathcal{O}(1/{M_p}^2)$ 
we derive $n_s={0.9577}$.
Therefore, it is essential for  the precise   $n_s$ calculation to include corrections up to $\mathcal{O}(1/{M_p}^4)$.


\section*{Acknowledgments}
The authors would like to thank Marcos A. G. Garcia for helpful discussions.
 This research work was supported by the Hellenic Foundation for Research and Innovation (H.F.R.I.) under the ``First Call for H.F.R.I. Research Projects to support Faculty members and Researchers and the procurement of high-cost research equipment grant'' (Project Number: 824).




\begin{thebibliography}{9} 

\bibitem{Abbott:2016blz}
B.~Abbott \textit{et al.} [LIGO Scientific and Virgo],
Phys. Rev. Lett. \textbf{116} (2016) no.6, 061102
doi:10.1103/PhysRevLett.116.061102
[arXiv:1602.03837 [gr-qc]].

\bibitem{Abbott:2016nmj}
B.~P.~Abbott \textit{et al.} [LIGO Scientific and Virgo],
Phys. Rev. Lett. \textbf{116} (2016) no.24, 241103
doi:10.1103/PhysRevLett.116.241103
[arXiv:1606.04855 [gr-qc]].



\bibitem{Abbott:2017oio}
B.~Abbott \textit{et al.} [LIGO Scientific and Virgo],
Phys. Rev. Lett. \textbf{119} (2017) no.14, 141101
doi:10.1103/PhysRevLett.119.141101
[arXiv:1709.09660 [gr-qc]].

\bibitem{Arzoumanian:2020vkk}
Z.~Arzoumanian \textit{et al.} [NANOGrav],
Astrophys. J. Lett. \textbf{905} (2020) no.2, L34
doi:10.3847/2041-8213/abd401
[arXiv:2009.04496 [astro-ph.HE]].

\bibitem{Arzoumanian:2018saf}
Z.~Arzoumanian \textit{et al.} [NANOGRAV],
Astrophys. J. \textbf{859} (2018) no.1, 47
doi:10.3847/1538-4357/aabd3b
[arXiv:1801.02617 [astro-ph.HE]].

\bibitem{Aggarwal:2018mgp}
K.~Aggarwal, Z.~Arzoumanian, P.~T.~Baker, A.~Brazier, M.~R.~Brinson, P.~R.~Brook, S.~Burke-Spolaor, S.~Chatterjee, J.~M.~Cordes and N.~J.~Cornish, \textit{et al.}
Astrophys. J. \textbf{880} (2019), 2
doi:10.3847/1538-4357/ab2236
[arXiv:1812.11585 [astro-ph.GA]].


\bibitem{Audley:2017drz}
P.~Amaro-Seoane \textit{et al.} [LISA],
[arXiv:1702.00786 [astro-ph.IM]].

\bibitem{Sato:2017dkf}
S.~Sato, S.~Kawamura, M.~Ando, T.~Nakamura, K.~Tsubono, A.~Araya, I.~Funaki, K.~Ioka, N.~Kanda and S.~Moriwaki, \textit{et al.}
J. Phys. Conf. Ser. \textbf{840} (2017) no.1, 012010
doi:10.1088/1742-6596/840/1/012010

\bibitem{Sathyaprakash:2009xs}
B.~S.~Sathyaprakash and B.~F.~Schutz,
Living Rev. Rel. \textbf{12} (2009), 2
doi:10.12942/lrr-2009-2
[arXiv:0903.0338 [gr-qc]].

\bibitem{Yagi:2011wg}
K.~Yagi and N.~Seto,
Phys. Rev. D \textbf{83} (2011), 044011
[erratum: Phys. Rev. D \textbf{95} (2017) no.10, 109901]
doi:10.1103/PhysRevD.83.044011
[arXiv:1101.3940 [astro-ph.CO]].


\bibitem{Luo:2015ght}
J.~Luo \textit{et al.} [TianQin],
Class. Quant. Grav. \textbf{33} (2016) no.3, 035010
doi:10.1088/0264-9381/33/3/035010
[arXiv:1512.02076 [astro-ph.IM]].



\bibitem{Zhao:2013bba}
W.~Zhao, Y.~Zhang, X.~P.~You and Z.~H.~Zhu,
Phys. Rev. D \textbf{87} (2013) no.12, 124012
doi:10.1103/PhysRevD.87.124012
[arXiv:1303.6718 [astro-ph.CO]].




\bibitem{Ballesteros:2017fsr}
G.~Ballesteros and M.~Taoso,
Phys. Rev. D \textbf{97} (2018) no.2, 023501
doi:10.1103/PhysRevD.97.023501
[arXiv:1709.05565 [hep-ph]].




\bibitem{Gao:2018pvq}
T.~Gao and Z.~Guo,
Phys. Rev. D \textbf{98} (2018) no.6, 063526
doi:10.1103/PhysRevD.98.063526
[arXiv:1806.09320 [hep-ph]].


\bibitem{Cicoli:2018asa}
M.~Cicoli, V.~A.~Diaz and F.~G.~Pedro,
JCAP \textbf{06} (2018), 034
doi:10.1088/1475-7516/2018/06/034
[arXiv:1803.02837 [hep-th]].

\bibitem{Dalianis:2018frf}
I.~Dalianis, A.~Kehagias and G.~Tringas,
JCAP \textbf{01} (2019), 037
doi:10.1088/1475-7516/2019/01/037
[arXiv:1805.09483 [astro-ph.CO]].


\bibitem{Garcia-Bellido:2017mdw}
J.~Garcia-Bellido and E.~Ruiz Morales,
Phys. Dark Univ. \textbf{18} (2017), 47-54
doi:10.1016/j.dark.2017.09.007
[arXiv:1702.03901 [astro-ph.CO]].



\bibitem{Ezquiaga:2017fvi}
J.~M.~Ezquiaga, J.~Garcia-Bellido and E.~Ruiz Morales,
Phys. Lett. B \textbf{776} (2018), 345-349
doi:10.1016/j.physletb.2017.11.039
[arXiv:1705.04861 [astro-ph.CO]].

\bibitem{Nanopoulos:2020nnh}
D.~V.~Nanopoulos, V.~C.~Spanos and I.~D.~Stamou,
Phys. Rev. D \textbf{102} (2020) no.8, 083536
doi:10.1103/PhysRevD.102.083536
[arXiv:2008.01457 [astro-ph.CO]].

\bibitem{Stamou:2021qdk}
I.~D.~Stamou,
Phys. Rev. D \textbf{103} (2021) no.8, 083512
doi:10.1103/PhysRevD.103.083512
[arXiv:2104.08654 [hep-ph]].

\bibitem{Hertzberg:2017dkh}
M.~P.~Hertzberg and M.~Yamada,
Phys. Rev. D \textbf{97} (2018) no.8, 083509
doi:10.1103/PhysRevD.97.083509
[arXiv:1712.09750 [astro-ph.CO]].

\bibitem{Ballesteros:2019hus}
G.~Ballesteros, J.~Rey and F.~Rompineve,
JCAP \textbf{06} (2020), 014
doi:10.1088/1475-7516/2020/06/014
[arXiv:1912.01638 [astro-ph.CO]].




\bibitem{Kefala:2020xsx}
K.~Kefala, G.~P.~Kodaxis, I.~D.~Stamou and N.~Tetradis,
[arXiv:2010.12483 [astro-ph.CO]].

\bibitem{Dalianis:2021iig}
I.~Dalianis, G.~P.~Kodaxis, I.~D.~Stamou, N.~Tetradis and A.~Tsigkas-Kouvelis,
[arXiv:2106.02467 [astro-ph.CO]].

\bibitem{Braglia:2020fms}
M.~Braglia, D.~K.~Hazra, L.~Sriramkumar and F.~Finelli,
JCAP \textbf{08} (2020), 025
doi:10.1088/1475-7516/2020/08/025
[arXiv:2004.00672 [astro-ph.CO]].

\bibitem{Braglia:2020eai}
M.~Braglia, D.~K.~Hazra, F.~Finelli, G.~F.~Smoot, L.~Sriramkumar and A.~A.~Starobinsky,
JCAP \textbf{08} (2020), 001
doi:10.1088/1475-7516/2020/08/001
[arXiv:2005.02895 [astro-ph.CO]].

\bibitem{Braglia:2020taf}
M.~Braglia, X.~Chen and D.~K.~Hazra,
JCAP \textbf{03} (2021), 005
doi:10.1088/1475-7516/2021/03/005
[arXiv:2012.05821 [astro-ph.CO]].

\bibitem{Palma:2020ejf}
G.~A.~Palma, S.~Sypsas and C.~Zenteno,
Phys. Rev. Lett. \textbf{125} (2020) no.12, 121301
doi:10.1103/PhysRevLett.125.121301
[arXiv:2004.06106 [astro-ph.CO]].

\bibitem{Fumagalli:2020adf}
J.~Fumagalli, S.~Renaux-Petel, J.~W.~Ronayne and L.~T.~Witkowski,
[arXiv:2004.08369 [hep-th]].

\bibitem{Gundhi:2020kzm}
A.~Gundhi, S.~V.~Ketov and C.~F.~Steinwachs,
Phys. Rev. D \textbf{103} (2021) no.8, 083518
doi:10.1103/PhysRevD.103.083518
[arXiv:2011.05999 [hep-th]].

\bibitem{Khlopov:2015oda}
M.~Khlopov,
Symmetry \textbf{7} (2015) no.2, 815-842
doi:10.3390/sym7020815
[arXiv:1505.08077 [astro-ph.CO]].

\bibitem{Ketov:2019mfc}
S.~V.~Ketov and M.~Y.~Khlopov,
Symmetry \textbf{11} (2019) no.4, 511
doi:10.3390/sym11040511

\bibitem{Mukherjee:2021ags}
S.~Mukherjee and J.~Silk,
doi:10.1093/mnras/stab1932
[arXiv:2105.11139 [gr-qc]].

\bibitem{Mukherjee:2021itf}
S.~Mukherjee, M.~S.~P.~Meinema and J.~Silk,
[arXiv:2107.02181 [astro-ph.CO]].

\bibitem{Vagnozzi:2020gtf}
S.~Vagnozzi,
Mon. Not. Roy. Astron. Soc. \textbf{502} (2021) no.1, L11-L15
doi:10.1093/mnrasl/slaa203
[arXiv:2009.13432 [astro-ph.CO]].

\bibitem{Pi:2017gih}
S.~Pi, Y.~l.~Zhang, Q.~G.~Huang and M.~Sasaki,
JCAP \textbf{05} (2018), 042
doi:10.1088/1475-7516/2018/05/042
[arXiv:1712.09896 [astro-ph.CO]].

\bibitem{Cai:2019bmk}
R.~G.~Cai, Z.~K.~Guo, J.~Liu, L.~Liu and X.~Y.~Yang,
JCAP \textbf{06} (2020), 013
doi:10.1088/1475-7516/2020/06/013
[arXiv:1912.10437 [astro-ph.CO]].

\bibitem{Kohri:2020qqd}
K.~Kohri and T.~Terada,
Phys. Lett. B \textbf{813} (2021), 136040
doi:10.1016/j.physletb.2020.136040
[arXiv:2009.11853 [astro-ph.CO]].


\bibitem{Ashoorioon:2019xqc}
A.~Ashoorioon, A.~Rostami and J.~T.~Firouzjaee,
JHEP \textbf{07} (2021), 087
doi:10.1007/JHEP07(2021)087
[arXiv:1912.13326 [astro-ph.CO]].

\bibitem{Ashoorioon:2020hln}
A.~Ashoorioon, A.~Rostami and J.~T.~Firouzjaee,
Phys. Rev. D \textbf{103} (2021), 123512
doi:10.1103/PhysRevD.103.123512
[arXiv:2012.02817 [astro-ph.CO]].
\bibitem{Adams:2001vc}
J.~A.~Adams, B.~Cresswell and R.~Easther,
Phys. Rev. D \textbf{64} (2001), 123514
doi:10.1103/PhysRevD.64.123514
[arXiv:astro-ph/0102236 [astro-ph]].

\bibitem{Hazra:2010ve}
D.~K.~Hazra, M.~Aich, R.~K.~Jain, L.~Sriramkumar and T.~Souradeep,
JCAP \textbf{10} (2010), 008
doi:10.1088/1475-7516/2010/10/008
[arXiv:1005.2175 [astro-ph.CO]].



\bibitem{Linde:1993cn}
A.~D.~Linde,
Phys. Rev. D \textbf{49} (1994), 748-754
doi:10.1103/PhysRevD.49.748
[arXiv:astro-ph/9307002 [astro-ph]].

\bibitem{Linde:1991km}
A.~D.~Linde,
Phys. Lett. B \textbf{259} (1991), 38-47
doi:10.1016/0370-2693(91)90130-I


\bibitem{Copeland:1994vg}
E.~J.~Copeland, A.~R.~Liddle, D.~H.~Lyth, E.~D.~Stewart and D.~Wands,
Phys. Rev. D \textbf{49} (1994), 6410-6433
doi:10.1103/PhysRevD.49.6410
[arXiv:astro-ph/9401011 [astro-ph]].


\bibitem{Mollerach:1993sy}
S.~Mollerach, S.~Matarrese and F.~Lucchin,
Phys. Rev. D \textbf{50} (1994), 4835-4841
doi:10.1103/PhysRevD.50.4835
[arXiv:astro-ph/9309054 [astro-ph]].

\bibitem{Linde:1997sj}
A.~D.~Linde and A.~Riotto,
Phys. Rev. D \textbf{56} (1997), R1841-R1844
doi:10.1103/PhysRevD.56.R1841
[arXiv:hep-ph/9703209 [hep-ph]].

\bibitem{Akrami:2018odb}
Y.~Akrami \textit{et al.} [Planck],
Astron. Astrophys. \textbf{641} (2020), A10
doi:10.1051/0004-6361/201833887
[arXiv:1807.06211 [astro-ph.CO]].

\bibitem{Ade:2015lrj}
P.~Ade \textit{et al.} [Planck],
Astron. Astrophys. \textbf{594} (2016), A20
doi:10.1051/0004-6361/201525898
[arXiv:1502.02114 [astro-ph.CO]].

\bibitem{Dvali:1994ms}
G.~R.~Dvali, Q.~Shafi and R.~K.~Schaefer,
Phys. Rev. Lett. \textbf{73} (1994), 1886-1889
doi:10.1103/PhysRevLett.73.1886
[arXiv:hep-ph/9406319 [hep-ph]].


\bibitem{Bastero-Gil:2006zpr}
M.~Bastero-Gil, S.~F.~King and Q.~Shafi,
Phys. Lett. B \textbf{651} (2007), 345-351
doi:10.1016/j.physletb.2006.06.085
[arXiv:hep-ph/0604198 [hep-ph]].


\bibitem{Dimopoulos:1997fv}
S.~Dimopoulos, G.~R.~Dvali and R.~Rattazzi,
Phys. Lett. B \textbf{410} (1997), 119-124
doi:10.1016/S0370-2693(97)00970-2
[arXiv:hep-ph/9705348 [hep-ph]].


\bibitem{Dvali:1997uq}
G.~R.~Dvali, G.~Lazarides and Q.~Shafi,
Phys. Lett. B \textbf{424} (1998), 259-264
doi:10.1016/S0370-2693(98)00145-2
[arXiv:hep-ph/9710314 [hep-ph]].



\bibitem{Panagiotakopoulos:1997if}
C.~Panagiotakopoulos and N.~Tetradis,
Phys. Rev. D \textbf{59} (1999), 083502
doi:10.1103/PhysRevD.59.083502
[arXiv:hep-ph/9710526 [hep-ph]].

\bibitem{Tetradis:1997kp}
N.~Tetradis,
Phys. Rev. D \textbf{57} (1998), 5997-6002
doi:10.1103/PhysRevD.57.5997
[arXiv:astro-ph/9707214 [astro-ph]].

\bibitem{Lazarides:1998zf}
G.~Lazarides and N.~Tetradis,
Phys. Rev. D \textbf{58} (1998), 123502
doi:10.1103/PhysRevD.58.123502
[arXiv:hep-ph/9802242 [hep-ph]].

\bibitem{Clesse:2010iz}
S.~Clesse,
Phys. Rev. D \textbf{83} (2011), 063518
doi:10.1103/PhysRevD.83.063518
[arXiv:1006.4522 [gr-qc]].

\bibitem{Pallis:2009pq}
C.~Pallis,
JCAP \textbf{04} (2009), 024
doi:10.1088/1475-7516/2009/04/024
[arXiv:0902.0334 [hep-ph]].

\bibitem{Armillis:2012bs}
R.~Armillis and C.~Pallis,
[arXiv:1211.4011 [hep-ph]].

\bibitem{Pallis:2013dxa}
C.~Pallis and Q.~Shafi,
Phys. Lett. B \textbf{725} (2013), 327-333
doi:10.1016/j.physletb.2013.07.029
[arXiv:1304.5202 [hep-ph]].


\bibitem{Pallis:2014xva}
C.~Pallis and Q.~Shafi,
Phys. Lett. B \textbf{736} (2014), 261-266
doi:10.1016/j.physletb.2014.07.031
[arXiv:1405.7645 [hep-ph]].

\bibitem{Dimopoulos:2016tzn}
K.~Dimopoulos and C.~Owen,
JCAP \textbf{10} (2016), 020
doi:10.1088/1475-7516/2016/10/020
[arXiv:1606.06677 [hep-ph]].

\bibitem{Kadota:2017dbz}
K.~Kadota, T.~Kobayashi and K.~Sumita,
JCAP \textbf{11} (2017), 033
doi:10.1088/1475-7516/2017/11/033
[arXiv:1707.00813 [hep-ph]].


\bibitem{Rehman:2009nq}
M.~U.~Rehman, Q.~Shafi and J.~R.~Wickman,
Phys. Lett. B \textbf{683} (2010), 191-195
doi:10.1016/j.physletb.2009.12.010
[arXiv:0908.3896 [hep-ph]].



\bibitem{Yamaguchi:2004tn}
M.~Yamaguchi and J.~Yokoyama,
Phys. Rev. D \textbf{70} (2004), 023513
doi:10.1103/PhysRevD.70.023513
[arXiv:hep-ph/0402282 [hep-ph]].

\bibitem{Wu:2016fzp}
L.~Wu, S.~Hu and T.~Li,
Eur. Phys. J. C \textbf{77} (2017) no.3, 168
doi:10.1140/epjc/s10052-017-4741-9
[arXiv:1605.00735 [hep-ph]].


\bibitem{Moursy:2020sit}
A.~Moursy,
JHEP \textbf{02} (2021), 230
doi:10.1007/JHEP02(2021)230
[arXiv:2009.14149 [hep-ph]].

\bibitem{Antusch:2009ef}
S.~Antusch, K.~Dutta and P.~M.~Kostka,
Phys. Lett. B \textbf{677} (2009), 221-225
doi:10.1016/j.physletb.2009.05.043
[arXiv:0902.2934 [hep-ph]].


\bibitem{Lyth:2011kj}
D.~H.~Lyth,
[arXiv:1107.1681 [astro-ph.CO]].

\bibitem{Clesse:2015wea}
S.~Clesse and J.~Garc\'\i{}a-Bellido,
Phys. Rev. D \textbf{92} (2015) no.2, 023524
doi:10.1103/PhysRevD.92.023524
[arXiv:1501.07565 [astro-ph.CO]].


\bibitem{Kanazawa:2000ea}
T.~Kanazawa, M.~Kawasaki and T.~Yanagida,
Phys. Lett. B \textbf{482} (2000), 174-182
doi:10.1016/S0370-2693(00)00499-8
[arXiv:hep-ph/0002236 [hep-ph]].

\bibitem{Choi:2021yxz}
K.~Y.~Choi, S.~b.~Kang and R.~N.~Raveendran,
JCAP \textbf{06} (2021), 054
doi:10.1088/1475-7516/2021/06/054
[arXiv:2102.02461 [astro-ph.CO]].

\bibitem{Garcia-Bellido:2007bcw}
J.~Garcia-Bellido and D.~G.~Figueroa,
Phys. Rev. Lett. \textbf{98} (2007), 061302
doi:10.1103/PhysRevLett.98.061302
[arXiv:astro-ph/0701014 [astro-ph]].



\bibitem{Civiletti:2011qg}
M.~Civiletti, M.~U.~Rehman, Q.~Shafi and J.~R.~Wickman,
Phys. Rev. D \textbf{84} (2011), 103505
doi:10.1103/PhysRevD.84.103505
[arXiv:1104.4143 [astro-ph.CO]].

\bibitem{Kawasaki:2012rw}
M.~Kawasaki, K.~Saikawa and N.~Takeda,
Phys. Rev. D \textbf{87} (2013) no.10, 103521
doi:10.1103/PhysRevD.87.103521
[arXiv:1208.4160 [astro-ph.CO]].

\bibitem{Lazarides:2020zof}
G.~Lazarides, M.~U.~Rehman, Q.~Shafi and F.~K.~Vardag,
Phys. Rev. D \textbf{103} (2021) no.3, 035033
doi:10.1103/PhysRevD.103.035033
[arXiv:2007.01474 [hep-ph]].

\bibitem{Lazarides:2015cda}
G.~Lazarides and C.~Panagiotakopoulos,
Phys. Rev. D \textbf{92} (2015) no.12, 123502
doi:10.1103/PhysRevD.92.123502
[arXiv:1505.04926 [hep-ph]].
\bibitem{Cai:2018dig}
R.~g.~Cai, S.~Pi and M.~Sasaki,
Phys. Rev. Lett. \textbf{122} (2019) no.20, 201101
doi:10.1103/PhysRevLett.122.201101
[arXiv:1810.11000 [astro-ph.CO]].

\bibitem{Cai:2019amo}
R.~G.~Cai, S.~Pi, S.~J.~Wang and X.~Y.~Yang,
JCAP \textbf{05} (2019), 013
doi:10.1088/1475-7516/2019/05/013
[arXiv:1901.10152 [astro-ph.CO]].


\bibitem{Domenech:2021wkk}
G.~Dom\`enech, V.~Takhistov and M.~Sasaki,
[arXiv:2105.06816 [astro-ph.CO]].
LaTeX (EU)

\bibitem{Pi:2020otn}
S.~Pi and M.~Sasaki,
JCAP \textbf{09} (2020), 037
doi:10.1088/1475-7516/2020/09/037
[arXiv:2005.12306 [gr-qc]].


\bibitem{Kodama:2011vs}
H.~Kodama, K.~Kohri and K.~Nakayama,
Prog. Theor. Phys. \textbf{126} (2011), 331-350
doi:10.1143/PTP.126.331
[arXiv:1102.5612 [astro-ph.CO]].

\bibitem{Ketov:2016gej}
S.~V.~Ketov and T.~Terada,
Eur. Phys. J. C \textbf{76} (2016) no.8, 438
doi:10.1140/epjc/s10052-016-4283-6
[arXiv:1606.02817 [hep-th]].

\bibitem{Cremmer1979}
E.~Cremmer, B.~Julia, J.~Scherk, S.~Ferrara, L.~Girardello and P.~van Nieuwenhuizen
Nuclear Physics B, Volume 147, Issues 1–2, 1979, Pages 105-131,
doi://doi.org/10.1016/0550-3213(79)90417-6.




\bibitem{Bernardo:2016jdr}
H.~Bernardo and H.~Nastase,
JHEP \textbf{09} (2016), 071
doi:10.1007/JHEP09(2016)071
[arXiv:1605.01934 [hep-th]].
\bibitem{Clesse:2013jra}
S.~Clesse, B.~Garbrecht and Y.~Zhu,
Phys. Rev. D \textbf{89} (2014) no.6, 063519
doi:10.1103/PhysRevD.89.063519
[arXiv:1304.7042 [astro-ph.CO]].


\bibitem{Sasaki:1995aw}
M.~Sasaki and E.~D.~Stewart,
Prog. Theor. Phys. \textbf{95} (1996), 71-78
doi:10.1143/PTP.95.71
[arXiv:astro-ph/9507001 [astro-ph]].

\bibitem{Fujita:2014tja}
T.~Fujita, M.~Kawasaki and Y.~Tada,
JCAP \textbf{10} (2014), 030
doi:10.1088/1475-7516/2014/10/030
[arXiv:1405.2187 [astro-ph.CO]].



\bibitem{Ringeval:2007am}
C.~Ringeval,
Lect. Notes Phys. \textbf{738} (2008), 243-273
doi:10.1007/978-3-540-74353-8\_7
[arXiv:astro-ph/0703486 [astro-ph]].

\bibitem{Lymanalpha}
S.~Bird, H.V. Peiris, M.~ Viel , L.~Verde 
MNRAS, May 2011,Vol 413, Issue 3, 1717--1728
doi:10.1111/j.1365-2966.2011.18245.x
[arXiv:1010.1519  [astro-ph.CO]].

\bibitem{Fixsen:1996nj}
D.~J.~Fixsen, E.~S.~Cheng, J.~M.~Gales, J.~C.~Mather, R.~A.~Shafer and E.~L.~Wright,
Astrophys. J. \textbf{473} (1996), 576
doi:10.1086/178173
[arXiv:astro-ph/9605054 [astro-ph]].

\bibitem{Nakama:2014vla}
T.~Nakama, T.~Suyama and J.~Yokoyama,
Phys. Rev. Lett. \textbf{113} (2014), 061302
doi:10.1103/PhysRevLett.113.061302
[arXiv:1403.5407 [astro-ph.CO]].


\bibitem{Jeong:2014gna}
D.~Jeong, J.~Pradler, J.~Chluba and M.~Kamionkowski,
Phys. Rev. Lett. \textbf{113} (2014), 061301
doi:10.1103/PhysRevLett.113.061301
[arXiv:1403.3697 [astro-ph.CO]].



\bibitem{Acquaviva:2002ud}
V.~Acquaviva, N.~Bartolo, S.~Matarrese and A.~Riotto,
Nucl. Phys. B \textbf{667} (2003), 119-148
doi:10.1016/S0550-3213(03)00550-9
[arXiv:astro-ph/0209156 [astro-ph]].

\bibitem{Mollerach:2003nq}
S.~Mollerach, D.~Harari and S.~Matarrese,
Phys. Rev. D \textbf{69} (2004), 063002
doi:10.1103/PhysRevD.69.063002
[arXiv:astro-ph/0310711 [astro-ph]].

\bibitem{Ananda:2006af}
K.~N.~Ananda, C.~Clarkson and D.~Wands,
Phys. Rev. D \textbf{75} (2007), 123518
doi:10.1103/PhysRevD.75.123518
[arXiv:gr-qc/0612013 [gr-qc]].

\bibitem{Baumann:2007zm}
D.~Baumann, P.~J.~Steinhardt, K.~Takahashi and K.~Ichiki,
Phys. Rev. D \textbf{76} (2007), 084019
doi:10.1103/PhysRevD.76.084019
[arXiv:hep-th/0703290 [hep-th]].



\bibitem{Espinosa:2018eve}
J.~R.~Espinosa, D.~Racco and A.~Riotto,
JCAP \textbf{09} (2018), 012
doi:10.1088/1475-7516/2018/09/012
[arXiv:1804.07732 [hep-ph]].

\bibitem{Kohri:2018awv}
K.~Kohri and T.~Terada,
Phys. Rev. D \textbf{97} (2018) no.12, 123532
doi:10.1103/PhysRevD.97.123532
[arXiv:1804.08577 [gr-qc]].

\bibitem{Matarrese:1997ay}
S.~Matarrese, S.~Mollerach and M.~Bruni,
Phys. Rev. D \textbf{58} (1998), 043504
doi:10.1103/PhysRevD.58.043504
[arXiv:astro-ph/9707278 [astro-ph]].


\bibitem{Maggiore:1999vm}
M.~Maggiore,
Phys. Rept. \textbf{331} (2000), 283-367
doi:10.1016/S0370-1573(99)00102-7
[arXiv:gr-qc/9909001 [gr-qc]].



\bibitem{DeLuca:2020agl}
V.~De Luca, G.~Franciolini and A.~Riotto,
Phys. Rev. Lett. \textbf{126} (2021) no.4, 041303
doi:10.1103/PhysRevLett.126.041303
[arXiv:2009.08268 [astro-ph.CO]].



\bibitem{Carr:1975qj}
B.~J.~Carr,
Astrophys. J. \textbf{201} (1975), 1-19
doi:10.1086/153853









\bibitem{Harada:2013epa}
T.~Harada, C.~M.~Yoo and K.~Kohri,
Phys. Rev. D \textbf{88} (2013) no.8, 084051
[erratum: Phys. Rev. D \textbf{89} (2014) no.2, 029903]
doi:10.1103/PhysRevD.88.084051
[arXiv:1309.4201 [astro-ph.CO]].



\bibitem{Musco:2008hv}
I.~Musco, J.~C.~Miller and A.~G.~Polnarev,
Class. Quant. Grav. \textbf{26} (2009), 235001
doi:10.1088/0264-9381/26/23/235001
[arXiv:0811.1452 [gr-qc]].

\bibitem{Musco:2004ak}
I.~Musco, J.~C.~Miller and L.~Rezzolla,
Class. Quant. Grav. \textbf{22} (2005), 1405-1424
doi:10.1088/0264-9381/22/7/013
[arXiv:gr-qc/0412063 [gr-qc]].

\bibitem{Musco:2012au}
I.~Musco and J.~C.~Miller,
Class. Quant. Grav. \textbf{30} (2013), 145009
doi:10.1088/0264-9381/30/14/145009
[arXiv:1201.2379 [gr-qc]].

\bibitem{Musco:2018rwt}
I.~Musco,
Phys. Rev. D \textbf{100} (2019) no.12, 123524
doi:10.1103/PhysRevD.100.123524
[arXiv:1809.02127 [gr-qc]].



\bibitem{Escriva:2019phb}
A.~Escriv\`a, C.~Germani and R.~K.~Sheth,
Phys. Rev. D \textbf{101} (2020) no.4, 044022
doi:10.1103/PhysRevD.101.044022
[arXiv:1907.13311 [gr-qc]].


\bibitem{Escriva:2020tak}
A.~Escriv\`a, C.~Germani and R.~K.~Sheth,
JCAP \textbf{01} (2021), 030
doi:10.1088/1475-7516/2021/01/030
[arXiv:2007.05564 [gr-qc]].


\bibitem{Musco:2020jjb}
I.~Musco, V.~De Luca, G.~Franciolini and A.~Riotto,
Phys. Rev. D \textbf{103} (2021) no.6, 063538
doi:10.1103/PhysRevD.103.063538
[arXiv:2011.03014 [astro-ph.CO]].



\bibitem{Carr:2009jm}
B.~J.~Carr, K.~Kohri, Y.~Sendouda and J.~Yokoyama,
Phys. Rev. D \textbf{81} (2010), 104019
doi:10.1103/PhysRevD.81.104019
[arXiv:0912.5297 [astro-ph.CO]].


\bibitem{Inoue:2017csr}
Y.~Inoue and A.~Kusenko,
JCAP \textbf{10} (2017), 034
doi:10.1088/1475-7516/2017/10/034
[arXiv:1705.00791 [astro-ph.CO]].




\bibitem{Montero-Camacho:2019jte}
P.~Montero-Camacho, X.~Fang, G.~Vasquez, M.~Silva and C.~M.~Hirata,
JCAP \textbf{08} (2019), 031
doi:10.1088/1475-7516/2019/08/031
[arXiv:1906.05950 [astro-ph.CO]].

\bibitem{Katz:2018zrn}
A.~Katz, J.~Kopp, S.~Sibiryakov and W.~Xue,
JCAP \textbf{12} (2018), 005
doi:10.1088/1475-7516/2018/12/005
[arXiv:1807.11495 [astro-ph.CO]].

\bibitem{Poulin:2017bwe}
V.~Poulin, P.~D.~Serpico, F.~Calore, S.~Clesse and K.~Kohri,
Phys. Rev. D \textbf{96} (2017) no.8, 083524
doi:10.1103/PhysRevD.96.083524
[arXiv:1707.04206 [astro-ph.CO]].

\bibitem{Capela:2013yf}
F.~Capela, M.~Pshirkov and P.~Tinyakov,
Phys. Rev. D \textbf{87} (2013) no.12, 123524
doi:10.1103/PhysRevD.87.123524
[arXiv:1301.4984 [astro-ph.CO]].

\bibitem{Niikura:2017zjd}
H.~Niikura, M.~Takada, N.~Yasuda, R.~H.~Lupton, T.~Sumi, S.~More, T.~Kurita, S.~Sugiyama, A.~More and M.~Oguri, \textit{et al.}
Nature Astron. \textbf{3} (2019) no.6, 524-534
doi:10.1038/s41550-019-0723-1
[arXiv:1701.02151 [astro-ph.CO]].


\bibitem{Wyrzykowski:2011tr}
L.~Wyrzykowski, J.~Skowron, S.~Kozlowski, A.~Udalski, M.~K.~Szymanski, M.~Kubiak, G.~Pietrzynski, I.~Soszynski, O.~Szewczyk and K.~Ulaczyk, \textit{et al.}
Mon. Not. Roy. Astron. Soc. \textbf{416} (2011), 2949
doi:10.1111/j.1365-2966.2011.19243.x
[arXiv:1106.2925 [astro-ph.GA]].

\bibitem{Griest:2013esa}
K.~Griest, A.~M.~Cieplak and M.~J.~Lehner,
Phys. Rev. Lett. \textbf{111} (2013) no.18, 181302
doi:10.1103/PhysRevLett.111.181302


\bibitem{Tisserand:2006zx}
P.~Tisserand \textit{et al.} [EROS-2],
Astron. Astrophys. \textbf{469} (2007), 387-404
doi:10.1051/0004-6361:20066017
[arXiv:astro-ph/0607207 [astro-ph]].

\bibitem{Ali-Haimoud:2016mbv}
Y.~Ali-Ha\"\i{}moud and M.~Kamionkowski,
Phys. Rev. D \textbf{95} (2017) no.4, 043534
doi:10.1103/PhysRevD.95.043534
[arXiv:1612.05644 [astro-ph.CO]].




\bibitem{Gaggero:2016dpq}
D.~Gaggero, G.~Bertone, F.~Calore, R.~M.~T.~Connors, M.~Lovell, S.~Markoff and E.~Storm,
Phys. Rev. Lett. \textbf{118} (2017) no.24, 241101
doi:10.1103/PhysRevLett.118.241101
[arXiv:1612.00457 [astro-ph.HE]].


\bibitem{Barbieri:1987fn}
R.~Barbieri and G.~F.~Giudice,
Nucl. Phys. B \textbf{306} (1988), 63-76
doi:10.1016/0550-3213(88)90171-X


\end{thebibliography}
\end{document}